\documentclass[acmsmall, review]{acmart}

\usepackage{graphicx,balance,multirow,xspace,subcaption,longtable}
\usepackage{tabularx,multirow,booktabs,blindtext}
\usepackage{makecell,amsmath,cleveref,rotating}
\usepackage[font={small}, textfont=md]{caption}
\usepackage[T1]{fontenc}
\usepackage[utf8]{inputenc}
\usepackage[draft,inline,nomargin,index]{fixme}
\usepackage{booktabs}
\usepackage{xcolor}

\PassOptionsToPackage{hyphens}{url}
\usepackage{hyperref} 
\newcolumntype{P}[1]{>{\arraybackslash}p{#1}}
\newcolumntype{X}[1]{>{\centering\arraybackslash}p{#1}}

\expandafter\def\expandafter\UrlBreaks\expandafter{\UrlBreaks%  save the current one
  \do\a\do\b\do\c\do\d\do\e\do\f\do\g\do\h\do\i\do\j%
  \do\k\do\l\do\m\do\n\do\o\do\p\do\q\do\r\do\s\do\t%
  \do\u\do\v\do\w\do\x\do\y\do\z\do\A\do\B\do\C\do\D%
  \do\E\do\F\do\G\do\H\do\I\do\J\do\K\do\L\do\M\do\N%
  \do\O\do\P\do\Q\do\R\do\S\do\T\do\U\do\V\do\W\do\X%
  \do\Y\do\Z}

\crefformat{section}{\S#2#1#3}
\crefformat{subsection}{\S#2#1#3}
\crefformat{subsubsection}{\S#2#1#3}

\fxsetup{theme=colorsig,mode=multiuser,inlineface=\itshape,envface=\itshape}
\FXRegisterAuthor{hh}{ahh}{Hussam}
\FXRegisterAuthor{rn}{arn}{Rishab}

\hypersetup{%
  pdftitle={Reddit Interventions},
  pdfauthor={},
  pdfkeywords={},
  bookmarksnumbered,
  pdfstartview={FitH},
  colorlinks,
  citecolor=black,
  filecolor=black,
  linkcolor=black,
  urlcolor=linkColor,
  breaklinks=true,
  hypertexnames=false
}

\newcommand\clearrow{\global\let\rowmac\relax}

\newcommand{\para}[1]{{\vspace{.05in} \bf \noindent #1 }}
\newcommand{\parait}[1]{{\vspace{.05in} \em \noindent #1 }}

\newcommand{\subreddit}[1]{\emph{r/#1}}

\renewcommand{\footnotesize}{\scriptsize}

\newcommand{\eg}{e.g.,\ }
\newcommand{\etal}{et al.\xspace}
\newcommand{\ie}{i.e.,\ }

\setcopyright{none}

\begin{document}

\renewcommand\footnotetextcopyrightpermission[1]{} % removes footnote with conference information in first column
\pagestyle{plain} % removes running headers

\renewcommand{\sectionautorefname}{\S}
\renewcommand{\subsectionautorefname}{\S}
\renewcommand{\subsubsectionautorefname}{\S}

\title{Making a Radical Misogynist}
\subtitle{How online social engagement with the Manosphere influences traits of radicalization}

\author{Hussam Habib}
\affiliation{%
  \institution{University of Iowa}
  \city{Iowa City}
  \country{USA}}
\email{hussam-habib@uiowa.edu}

\author{Padmini Srinivasan}
\affiliation{%
  \institution{University of Iowa}
  \city{Iowa City}
  \country{USA}}
\email{padmini-srinivasan@uiowa.edu}

\author{Rishab Nithyanand}
\affiliation{%
  \institution{University of Iowa}
  \city{Iowa City}
  \country{USA}}
\email{rishab-nithyanand@uiowa.edu}

\renewcommand{\shortauthors}{H. Habib \etal}
\begin{abstract}

  The algorithms and the interactions facilitated by online platforms have been
  used by radical groups to recruit vulnerable individuals to their cause. This
  has resulted in the sharp growth of violent events and deteriorating online
  discourse. The Manosphere, a collection of radical anti-feminist communities,
  is one such group which has attracted attention due to their rapid growth and
  increasingly violent real world outbursts. 
  In this paper, we examine the social engagements between Reddit users who have
  participated in feminist discourse and the Manosphere communities on Reddit
  to understand the process of development of traits associated with the
  adoption of extremist ideologies. 
  By using existing research on the psychology of radicalization we track how
  specific types of social engagement with the Manosphere influence the
  development of traits associated with radicalization. 
  Our findings show that: (1) participation, even by the simple act of joining the
  Manosphere, has a significant influence on the language and outlook traits of
  a user, (2) Manosphere elites are extremely effective propagators of radical
  traits and cause their increase even outside the Manosphere, and (3)
  community perception can heavily influence a user's behavior.
Finally, we examine
  how our findings can help draft community and   platform moderation policies
  to help mitigate the problem of online   radicalization.

\end{abstract}

\sloppy

\maketitle

\section{Introduction}

% Introductory para: The hook
% 
\para{We need to understand how extremist ideologies are adopted.}
Communities hosted by online platforms have now become a core mechanism for
recruiting and organizing vulnerable individuals into extremist groups. The
impact of such recruitment has resulted in harms extending beyond online
discourse. For example, investigations have uncovered that violent events such
as the January $6^{\text{th}}$ attack on the US Capitol
\cite{CapitolRiotsOnline}, the 2017 Unite the Right rally in Charlottesville
\cite{UTROnline}, and the Alek Minassian attack in Toronto \cite{AMOnline} were
either planned by or celebrated in radical online communities hosted on Reddit,
Facebook, 4chan, and other platforms. 
The use of online platforms for recruitment into extremist groups is not new
--- Facebook, YouTube, and Twitter were used as tools for recruitment into
Islamic extremist groups as early as 2010 \cite{Weimann-BJWA2010}. What is new,
however, is the sharp rise in domestic extremism within the United States
resulting from organization in online communities. In fact, the US Federal
Bureau of Investigation (FBI) now lists social media and online radicalization
as one of the largest factors contributing to the rise of domestic terrorism
\cite{TerrorismFBI} and a 2021 joint report by the US Department of Homeland
Security and the FBI list ``lone offenders who are often radicalized online''
as the ``greatest terrorism threat to the homeland'' \cite{jointreport}.
In order to mitigate these threats, it is paramount that we understand how
online platforms and discourse can be used to influence participants into
adopting extremist ideologies. 

\para{The recent growth of online misogynist extremism is alarming.} 
Amongst the newly emerging extremist groups, the online ``anti-feminist''
movement has been prominent. With the rise and adoption of feminism in the
mainstream, researchers have observed and documented the development of an
``anti-feminist'' communities in online platforms. Although focused largely on
the social issues faced by young men in modern society during its inception, in
the mid-2010s (roughly coinciding with the coverage of Gamergate
\cite{massanari2017gamergate}),
ethnographers noted a distinct ideological shift in many of these communities
towards anti-feminism,  misogyny, and male supremacism. Along with encouraging
and celebrating violent outbursts towards women, many of these online
communities also promote extreme political ideologies including state mandated
sexual partners and removal of voting rights for women. In recent
years, researchers have named this conglomerate of online communities ``the
Manosphere'' \cite{farrell2019exploring,marwick2018drinking}. Despite being
frequently referred to as a collective, the Manosphere comprises of four
distinct ideological subgroups: the involuntary celibates who blame women and
the rise of feminism for their low social status \cite{OMalley-JIV2020}, the
separatists (\eg Men Go Their Own Way and voluntary celibates) who believe in
a gynocentric conspiracy that society is corrupted by women with no possibility
of collective salvation and therefore pursue the goal of complete separation
from all women \cite{wright2020pussy}, the seduction strategists (\eg Pick Up
Artists and The Red Pill) which promotes the objectification of women and has
been identified as a gateway to acceptance of rape culture \cite{news-pua-rape,
news-pua-shooter, van2021digesting}, and the men's rights activists (MRA) that
generally views social systems as a zero-sum game and therefore portrays
feminist or pro-equality programs and policies as harmful to men's rights
\cite{splc-mensrights}. In recent years, there have been numerous acts of
violence towards women that have emerged from members of these online
communities \cite{ManosphereViolence}.

% The state-of-the-art: What do we already know so far
%
\para{The role of online social interactions in the development of extremist
ideologies remains largely unknown.}
The importance of understanding the factors involved in online radicalization
has not gone unnoticed. Extensive computational and social science research has
been ongoing to understand the role of online platforms' algorithms in
promoting and spreading extremist content, propaganda, and mis-/dis-information
related to topics popular in current American political discourse such as \eg
vaccines \cite{tang2021down}, immigration \cite{roose2019making,
faris2017partisanship}, and conspiracy theories \cite{bruns2021coronavirus}.
However, there is less contemporary 
research that has analyzed online user-to-user social engagements and their role
in the adoption of extremist ideologies. Existing research on the role of
user-to-user engagement has largely focused on identifying how Islamic
extremist groups used social media platforms to recruit vulnerable individuals
\cite{koehler2014radical, kadivar2017online, blaker2015islamic} and it remains
unclear if this research is directly applicable to
the social issues currently at the forefront of US domestic politics such as
anti-feminist extremism. Further, the dynamics of online social media, its
algorithms, and online discourse have changed significantly in the past decade
since much of this research was conducted. 

\para{Our contributions.}
In this paper, we focus on understanding process of adoption of ideological
extremism associated with the Manosphere. We do so by analyzing a variety of 
user-to-user engagements with the Manosphere. Then, through engaging with existing
psychology and threat assessment research (\Cref{sec:background}), we study the
relationships between specific types of engagements and the warning signs of
extremist behavior. 
This allows us to understand how different types of social engagements with the
Manosphere result in the development of traits associated with radical and
extremist behavior. More specifically, we ask the following research questions:

\begin{itemize}
  \item \textit{RQ1. How does participation in the Manosphere influence the
    traits associated with radicalization? (\Cref{sec:participation})} In our
    analysis, we consider three types of participation engagements: (1) joining
    a Manosphere community, (2) repeated submissions to the Manosphere, and (3)
    epistemic participation bubbles. We then use a control-treatment
    comparative analysis and a regression analysis to uncover how each of these
    types of participation engagement influence the warning signs associated
    with extremist behavior. Our results show a significant increase in
    language- and outlook-based warning behaviors even after a single
    participation event occurs.
  \item \textit{RQ2. How does interaction with elite Manosphere members outside
    the Manosphere influence the traits associated with radicalization?
    (\Cref{sec:interaction})}
    To answer this question, we consider two types of interactions that occur
    outside the Manosphere: (1)
    solicitation interactions where a user initiates an interaction with the
    elite and (2) a recruitment interaction where the elite initiates the
    interaction with a user. Our analysis shows that both forms of interaction
    have a heavy influence on the user's warning behaviors. Our findings bring
    to light the key differences in the operation of recruitment and
    solicitation interactions and the effects they have on users.
  \item \textit{RQ3. How does social status within and outside the Manosphere
    influence traits associated with radicalization? (\Cref{sec:perception})}
    We use Reddit's voting mechanism to develop proxies for social acceptance
    within the Manosphere and social rejection outside the Manosphere. We then
    use these proxies to uncover their influence on the accepted and rejected
    user's warning behaviors. We find that social acceptance within the
    Manosphere can have a sharp influence on a user's warning behaviors ---
    even after a single occurrence.
    
\end{itemize}

Taken together, our investigation yields insights into the specific influences
of different forms of user engagement with the Manosphere on the personality
traits commonly associated with radicalization and extremist behavior. These
insights can improve online moderation and content policy standards to mitigate
the threat of misogynist and other forms of extremism (\Cref{sec:discussion}).

%What are the radical pathways in social media platforms? To answer this
%question, we focus on social elements facilitated by the design of social media
%platforms and attempt to identify any radical pathways they enable. The online
%social elements we monitor are communities, users, and scores. These elements
%were designed to make online experiences more engaging and are fundamental to
%many social media platforms such as Reddit. However, the same elements have been
%seen to be exploited by radicalized groups to spread their agenda and grievances
%in attempts to indoctrinate and radicalize individuals. In our work, in order to
%understand how users radicalize, we attempt to identify the extent to which
%these elements influence user behavior and uncover how these elements are
%exploited to radicalize individuals. Since governance and intervention
%strategies by Reddit can moderate these elements, by understanding the role of
%these elements in radicalizing individuals we can better determine ways to block
%radical pathways. In our work, we focus on the process of misogynistic
%radicalization and measure how these elements influence warning behaviors of
%individuals.

\section{Background: Warning Behaviors and Traits of Radicalization}
\label{sec:background}

Radicalization is a complex social process of an individual adopting extreme
ideologies with potentially violent outcomes \cite{gaspar2020radicalization,
reidy2018radicalization}. Over the past decade, psychology and threat
assessment researchers have been studying the behaviors and traits exhibited by
extremists in order to better understand their progression and development.
The traits common in radicalized individuals are commonly referred to as
\emph{warning behaviors}. Although these warning behaviors were originally used
to identify and construct theoretical models of \emph{offline} radicalization
\cite{smith2018radicalization, gill2015lone, meloywarning,
elster1996rationality, baele2017lone, doosje2013determinants}, more recent
research has validated specific text analysis methods for their identification
in the \emph{online} context \cite{grover2019detecting, cohen2014detecting,
kaati2016linguistic}. 
%
% \para{Threat assessment toolkits.} 
Following the identification of several warning behaviors and an initial
understanding of the radicalization process, researchers have developed threat
assessment toolkits. These toolkits are essentially curated lists of specific
warning behaviors that help identify, in the clinical setting, individuals who
are ideologically radicalized or in the process of ideological radicalization. 
These toolkits differ from their `general violence' identification toolkit
counterparts in that they specifically probe risk factors associated with
ideologies, morality, and grievances \cite{assessments-dhs2018} which are
largely ignored in general violence toolkits. In our work, we focus on
understanding how social engagements influence eight specific traits or
warning behaviors, common to the following toolkits:

\begin{itemize}
  \item {TRAP-18}. The Terrorist and Radicalization Assessment Protocol
    proposed by Meloy \cite{meloy2018operational} identifies eight proximal and
    ten long-term distil warning behaviors to assess radicalization in
    individuals. Our study uses already validated text analysis to obtain
    measures of several proximal warning behaviors identified in TRAP-18. 
  \item {VERA-2}. The Violent Extremism Risk Assessment proposed by
    Pressman \cite{pressman2012calibrating} includes 25 warning behaviors that
    may be used to assess risk of radicalization for political ideologies.
    These warning behaviors are categorized into beliefs and attitudes, intent
    and contextual factors, historical factors, protective and social factors,
    and commitment and motivation factors. In our study, since we rely
    exclusively on user submitted text, we are only able to analyze warning
    behaviors identified in VERA's `beliefs and attitude' and `commitment and
    motivation' factors.
  \item {ERG-22+}. The Extremist Risk Guidelines proposed by Lloyd and Dean
    \cite{lloyd2019extremist, lloyd2015development} identifies 22 warning
    behaviors distributed in three groups: beliefs and motivations, intent, and
    capability. The ERG-22+ is recommended for use to specifically assess an
    individuals risk of committing violence on behalf of an extremist cause or
    group. Our analysis is focused on the `belief and motivation' warning
    behaviors.
  \item {IVP}. The Identifying Vulnerable Persons toolkit developed by Cole,
    \etal \cite{cole2010guidance} identifies 16 warning behaviors categorized
    into isolation factors, rhetoric, risk taking behavior, and contextual
    factors. The IVP is meant to assess an individual's vulnerability to
    recruitment into extremist groups. We limit our analysis to the rhetoric
    factors identified in the IVP.
\end{itemize}

\begin{table}[]
    \centering
    \resizebox{\textwidth}{!}{%
    \begin{tabular}{@{}lllll@{}}
    \toprule
    \textbf{Type} & \textbf{Trait} & \textbf{Definition} & \textbf{Measure} & \textbf{References} \\ \midrule
    \multirow{3}{*}{\textbf{Language}} 
      & Fixation & Increased preoccupation with a topic. & Frequency of occurrence of feminism keywords 
        & \cite{meloy2018operational, meloywarning, cohen2014detecting, grover2019detecting}     \\
      & Grievance & Expression of real of perceived injustice. & Grievance keywords \cite{van2021grievance} 
        & \cite{smith2018risk,doosje2013determinants,smith2018radicalization,mccauley2008mechanisms,corner2018mental} \\
     & Power & Increased need for power and authority. & LIWC `Power' keywords 
        & \cite{kaati2016linguistic,smith2004words,baele2017lone,chung2011using,smith2018risk} \\
     \midrule
    \multirow{1}{*}{\textbf{Emotions}} 
      & Anger & Exhibition of anger and aggression. & LIWC `Anger' keywords 
        & \cite{baele2017lone,meloy2018operational,smith2018risk,pressman2012calibrating,pressman2019internet} \\
     \midrule
    \multirow{2}{*}{\textbf{Outlook}} 
      & Negative & Negative outlook and sentiment. & \% of negative content (VADER) 
        & \cite{baele2017lone,stankov2010contemporary,douglas1999assessing} \\
      & Toxicity & Increase in hate speech and toxicity. & \% of toxic content (Perspective) 
        & \cite{sarma2017risk,egan2016can,pressman2012calibrating,pressman2019internet} \\
     \midrule
    \multirow{2}{*}{\textbf{Identification}} 
      & Ingroup & Increased identification with a group. & LIWC `We' keywords 
        & \cite{smith2018risk,cohen2014detecting,meloy2018operational,doosje2013determinants,sarma2017risk,mccauley2008mechanisms,berger2017extremist} \\
      & Outgroup & Increased mention of the outgroup. & LIWC `They' keywords 
        &
	\cite{smith2018risk,doosje2013determinants,sarma2017risk,pressman2012calibrating,berger2017extremist,scrivens2021examining} \\
     \bottomrule
    \end{tabular}%
    }
    \caption{A summary of the warning behaviors or traits studied in this
    paper, our methods for measuring them, and previous work which motivates
    their inclusion in our analysis.}
    \label{tab:background:traits}
\end{table}
 
\subsection{Measured warning behaviors and traits.} 
\label{sec:background:traits}
In our work, we focus on eight warning behaviors or traits which have been
repeatedly identified in the above toolkits. These eight warning behaviors were
selected because they were measurable from user submitted text using validated
text analysis methods. These are summarized in \Cref{tab:background:traits} and
described below. 

\para{{\it Caveat emptor.}} 
\emph{It is important to note that our work does not use these warning
behaviors to identify radicalized individuals. Rather, we seek to understand
what types of social engagements in the Manosphere influences changes,
particularly increases, in the warning behaviors of radicalization.}

\para{Fixation.}
% what is fixation
Cohen describes fixation as the `increasingly pathological preoccupation with
a person or a cause' \cite{cohen2014detecting}. Fixation towards a specific person,
cause or topic has been repeatedly identified as a key trait of individuals who
are vulnerable to or already radicalized. According to Meloy \etal
\cite{meloywarning}, a fixated person expresses increased preoccupation towards
a specific topic, having strident opinions about the topic along with levels of
disproportionate preoccupation that could lead to social and occupational
deterioration.
% fixation and misogynistic radicalization
% For our analysis, an increase in fixation towards feminism over the months would
% suggest a progress in the misogynistic radicalization of the individual.
% Fixation's relationship with radiclaization is even more visible in the process
% of misogynistic radicalization. The ideologies and beliefs present in the
% manosphere communities are driven more by the hatred towards women and feminism
% than they are driven by the discussion of men's rights. The recent most popular
% online event, GamerGate, was similarly fueled by the opposition of feminism in
% video game culture.  Therefore, increasing fixation on women and feminism are
% signs of misogynistic radicalization.
% 
% how we measure

\para{{\it Measuring fixation.}} We measure fixation as the percentage of text
submissions (comments or posts) made by a user that make references to
a pre-determined list of keywords associated with feminism. Such an approach
has been used in prior literature to characterize alt-right communities on
Reddit \cite{grover2019detecting}. We calculate this
metric based on text submissions made by a user each month for each user for
a period of 68 months to identify changes in fixation associated with feminism
and related topics. A consistent and significant increase in the monthly
fixation scores measured for a user would suggest an increasing fixation with
feminism. We refer the reader to \Cref{sec:methods} for detailed methodologies
related to construction of our keyword list and user selection.

% mention of a 
% particular topic in an online environment. To this end, we use our feminism
% keyword list \Fkeywords constructed with keywords representing discussion about
% feminism to identify submissions mentioning or discussing feminsim. We consider
% any submission containing at least one keyword from \Fkeywords as being related to
% fixaition. For a user, their fixation towards feminism for a particular month is
% represented by the percentage of their submissions contaning at least on keyword
% from \Fkeywords. We measure the fixation towards feminism for all \Fusers for all of
% the 68 months of our analysis. \note{provide statistics about feminism}.
% 

\para{Grievance.}
% what is grievance
Grievances (and perceived grievances) are defined as the feeling of injustice,
unfair treatment and frustration over suffering. Grievances can be real or
perceived. Perceived grievances has been repeatedly noted as a warning behavior
in threat assessment toolkits and research on lone wolf terrorists and
terrorist groups show that acts of violence are often fueled by perceived
grievances and injustices towards themselves or their groups
\cite{corner2018mental, mccauley2008mechanisms}. 
Prior research has also shown that white supremacist extremists use tactics to
induce perceived grievances (\eg by promoting the white replacement conspiracy
theory) to recruit and radicalize individuals \cite{berbrier2000victim}.
Similar tactics have been identified by social scientists in the context of the
Manosphere \cite{marwick2018drinking}.
% Exploration of the manosphere show multiple communities portraying themselves
% and men as the victims of radical feminism. Community position themselves as the
% defenders of masculinity and opposing feminism and any programs made for women.
% Studying the vocabulary of the Manosphere, Marwick shows how the manosphere
% adopts a defensible position of playing the victim to further their
% anti-feminist agenda by re-framing feminism as oppressive
% \cite{marwick2018drinknin}. These grievances take the shape of extreme
% ideologies by developing gyocentric conspiracies that execute cordinated attacks
% on masculinity. On occasions these grievances has led to violent outbursts from
% the members of these communities against women. By measuring the greivances of a
% user for a month, we expect to measure their progress in misogynistic
% radicalization.

% how we measure grievance
\para{{\it Measuring grievance.}}
We measure instances of communication fueled with grievances. To identify these
instances we use the grievance dictionary developed by van der Vegt \etal
\cite{van2021grievance}. This psycholinguistic 
dictionary was constructed by capturing vocabulary exhibiting psychological
and social concepts of grievance \cite{scrivens2018searching}. For each user,
we measure their average grievance score for each of the 68 months in our
study. The grievance score for a single text submission is the fraction of all
words in that submission that belong in the grievance dictionary. Observing
consistent and significant increases in a users' average grievance score for
a month would suggest an increase in the grievance trait for the user.

\para{Power.}
Power relates to the need for having impact, control, and influence over others
\cite{wilson2018comparative}. Previous research conducting linguistic analysis
of lone-actor manifestos show desire for power, identified by framing of
societal issues in the `leader-follower' and `authority' language, as a driving
force for their actions \cite{kaati2016linguistic}. This is further supported
by research findings that find motives of power and authority are 
the major driving forces for many terrorist groups and extremists
\cite{smith2018risk}.

\para{{\it Measuring power.}} We measure instances of communication containing
language associated with power. To identify these instances, we use the LIWC
(Linguistic Inquiry and Word Count \cite{tausczik2010psychological}) toolkit to
identify the levels of language associated with power in a user's text
submissions.
LIWC is a widely used and validated linguistic toolkit containing 80
dictionaries, each capturing vocabulary exhibiting different psychological,
emotional, and social concepts. One of these dictionaries lists words
associated with the psychological process of `Power'. This dictionary has been
used in previous linguistic analyses of lone-wolf terrorist manifestos
\cite{kaati2016linguistic}. For each user, we measure
their average power score for each of the 68 months in our study. The power
score for a single text submission is the fraction of all words in that
submission that are contained in LIWC's `power' word list. Observing
a consistent and significant increase in a users' average power score would
indicate an increase in the power trait for the user.

\para{Anger.}
In exploring the rationality of emotions, Elster \cite{elster1996rationality}
defined the action tendency of anger as being to strike. Research into
terrorist groups, lone-actor terrorists, threat assessment and psychological
research toolkits have all repeatedly shown high amounts of anger and
aggression, not necessarily targeted towards the perceived wrong-doer, as
a common trait amongst radicalized individuals.

\para{{\it Measuring anger.}} 
Similar to power, we use LIWC's `anger' words to measure anger expressed by
a user. We measure the anger score for a single text submission as the fraction
of all words that belong in LIWC's anger words. We then calculate and track
each users' average anger score for each of our 68 months. Observing
a consistent and significant increase in this score would indicate an increase
in the anger trait for the user.

\para{Negative outlook.}
A negative outlook of life and their current situation by an individual belies
the sense of lack of control over their own life. Increased expression of
negative sentiment is noted as a trait associated with individuals on the
pathways to radicalization in all our threat assessment toolkits. Prior social
science research also has shown how radical ideas and conspiracy theories
attempt to maintain higher negative sentiment and use them to incite more
negative outcomes. 
Furthermore, research into the Manosphere have noted the prevalence of negative
sentiment in their ideologies and beliefs. Negative self-images and a general
negative outlook on life are expressed commonly in the Incel communities in the
Manosphere.

\para{{\it Measuring negative outlook.}}
In order to measure negative outlook for an individual, we perform sentiment
analysis on their authored posts and comments. Using VADER
\cite{hutto2014vader}, we compute the sentiment associated with each text
submission. We then count the text submission as suggesting a negative outlook
if the compound score returned by VADER is less than ($-.05$), as recommended by
the authors of VADER. For each user and each month in our study, we track the
fraction of all the users' submissions that were found to have a negative
outlook. We expect that observing a consistent and significant increasing
fraction of negative sentiment submissions corresponds to an increase in the
users' negative outlook.

\para{Toxicity.}
% what is toxicity
Hate speech and toxic language is often used to dehumanize and express anger
towards the `opposition'. Researchers have observed radicalized and terrorist
groups often dehumanize their targets to justify violence and aggression towards
them \cite{wahlstrom2020dynamics, piazza2020politician}. All our threat
assessment toolkits include hate rhetoric as a warning behavior used to
identify extremists \cite{egan2016can}.
% toxicity and the manosphere
Notably, research on the Manosphere and its communities has been repeatedly
noted the dehumanization and objectification of women by the means of toxic
language \cite{krendel2020men}. 

\para{{\it Measuring toxicity.}}
% how we compute toxicity
For our analysis, we measure the percentage of toxic submissions made by user
each month. We identify toxic submissions using Google's Perspective API
\cite{Perspective-API-reddit}. Similar to uses of Perspective API in prior
work validating the API for Reddit, submissions scored with a probability of
0.5 or higher of being perceived toxic are labeled as toxic
\cite{pavlopoulos2019convai, obadimu2021developing}. For each user and each
month, we compute the fraction of all submissions made that month that are
classified as toxic. A significant increase in this metric is taken to
correspond with the users' toxicity trait.

\para{In- and out-group identification.}
% What is group identification
Group identification is a socio-psychological concept representing the
individual's sense of belonging with a group and their identity being defined
by this group membership. In the context of radicalization, group
identification aims to exploit social tribalism by generating `Us vs Them'
situations. Radicalizing individual generate a sense of in-group loyalty and
positive outlook towards in-group while also being hostile and negative to the
out-group. 
% Group identification in the manosphere
Research into the Manosphere has shown that it's members have a shared identity
and have constructed a major out-group (women) as the causes for their
grievances \cite{jaki2019online} and targets for violence \cite{bou2014gender}.

\para{{\it Measuring group identification.}}
% How we measure group identification
For our analysis, we measure both in-group identification and out-group
construction in online environment. Similar to prior works measuring in-group
and out-group scores \cite{cohen2014detecting,grover2019detecting}, we use the
psycho-linguistic dictionary LIWC.  
We use LIWC's first person plural pronouns category (`we') to measure in-group
identification. The use of first person plural pronouns in online discussion
represents the in-group identification of an individual. For each month for each
user, we aggregate all their submissions and compute the percentage of first
person plural pronouns present. This percentage represents their in-group
identification score. Increase in first person plural pronouns represents the
increased identification with the group.
Similarly, we use LIWC's third person plural pronouns category (`they') to
measure out-group construction. For each user, for each month we compute the
percentage of third person plural pronouns present in all of their submissions.
This percentage represents the out-group construction score. Increase in the use
of third person plural pronouns in online environment represents the
construction of an out-group.

\section{Data and Methods}\label{sec:methods}

In this section, we provide a description of our Reddit dataset
(\Cref{sec:methods:data}), our methods for identifying users of interest to
our study (\Cref{sec:methods:users}) and communities belonging to the Manosphere
(\Cref{sec:methods:subreddits}), and the influence of social engagement with
the Manosphere on the traits associated with radicalization
(\Cref{sec:methods:relationships}).

\subsection{The Reddit dataset}\label{sec:methods:data}

In this work, we specifically focus on Reddit users and their adoption of
anti-feminist ideologies. Reddit is essentially a discussion board consisting
of a variety of topical communities called \emph{subreddits}. Reddit users may
create their own subreddits or participate in existing ones by submitting
posts, commenting on posts or other user comments, or up-/down- voting comments
and posts made by other users. 
In total, Reddit has over 52M active daily users, over 1.2M subreddits
containing over 13B user submitted comments and posts, and is the
$6^{\text{th}}$ most popular site in the United States. 
We select Reddit as the subject of our study for three main reasons. First,
Reddit has been at the forefront of discussions surrounding the Manosphere
\cite{massanari2017gamergate, wright2020pussy, ribero2021evolution,
mamie2021anti} and other radical communities \cite{habib2019act,
habib2021proactive, grover2019detecting, hine2017kek, zannettou2017web}.
Second, unlike other social media platforms, Reddit has been welcoming and
responsive to academic research \cite{We-are-the-nerds}. Finally, Reddit data
is conveniently available through the Pushshift API \cite{PushShift-Reddit}.
We use the Pushshift Reddit dataset to gather all user submitted comments and
posts, along with their attached metadata for a 68 month period between 01/2015
and 08/2020 as the basis for our analysis.

\subsection{Identifying users of interest} \label{sec:methods:users}
Our research is specifically focused on understanding the influence of social
engagement with the Manosphere on anti-feminist radicalization. We begin by
identifying all Reddit users who were heavily engaged in discourse, via posts
or comments, on topics centered around feminism. We then study the engagements
between the Manosphere and these users to understand how warning behaviors are
influenced.
At a high-level, we identify participants in feminist discourse by constructing
a list of keywords associated with feminism and then identifying all users who
made any post or comment containing any of these keywords. 
This approach provides us with a dataset containing most users who have ever
engaged with feminism with a high recall rate of users highly engaged with
feminism --- regardless of the valence of the discourse. We then filter out
users who did not meet the minimum threshold for engagement with feminism.

\para{Defining participation in feminist discourse.} 
Our goal is to first identify all the users who engaged in any discussions
related to feminism using only the textual content of their comments and posts.
For our purposes, a user comment or post is defined as related to feminism if
any of our pre-identified feminism keywords are contained in it. A user is said
to have engaged in feminist discourse if any of their comments or posts are
related to feminism. 
Note that we do not restrict the subreddits in which this discourse may have
occurred. Therefore, we identify users who were engaged in feminist discourse
even in general communities such as \subreddit{politics}, \subreddit{news}, and
\subreddit{art}. 
This provides two benefits: (1) we are able to understand how engagements with
Manosphere members that occur outside a Manosphere community may influence
users' traits and (2) we are able to identify communities which serve as
a gateway to the Manosphere.

\para{{Curating a list of keywords related to feminist discourse.}}
As mentioned above, we utilize a list of keywords to identify comments and
posts submission engaging with feminism. 
To create this list of keywords, we begin with a manually curated seed list of
48 keywords and phrases (and their variants) sourced by the authors from
articles, communities, news, and books related to feminism and associated
topics. We do not distinguish between pro- and anti-feminist vocabularies. 
Our goal in creating a seed list is \emph{not} to capture a comprehensive list
of keywords and phrases. Instead, we aim to capture a sample of the keywords
and phrases associated with a well-rounded set of topics common to feminist
discourse.
Examples of words in this list include terms used in the discussion of
sexuality and objectification (\eg \emph{me too}, \emph{slut}), careers (\eg
\emph{girl boss}, \emph{equal pay}, \emph{mansplain}), relationships (\eg
\emph{hypergamy}, \emph{redpill}), reproductive rights (\eg \emph{abortion},
\emph{prolife}, \emph{prochoice}), and feminism (\eg \emph{feminazi},
\emph{patriarchy}).
We then use a snowballing method to grow our initial list of keywords and
phrases. First, we use word2vec \cite{mikolov2013efficient} to construct word
embedding over all unigrams, bigrams, and trigrams in a random 10\% sample of
all the Reddit comments and posts in our dataset. A sample was necessary due to
the computational limitations associated with processing billions of comments
and posts. Next, for each n-gram in our seed list, we use the constructed word
embedding to identify the 20 most similar, by meaning and context, n-grams.
Each of these ten n-grams is then manually validated and added to our list.
Finally, for inclusion in the final list of n-grams, we include all
conjugations and variants of all n-grams contained in our list. Therefore,
each entry in our list is manually validated. Our final list contains
158 n-grams and their variants.

\para{Identifying feminist discourse on Reddit.}
We begin by selecting all Reddit comments and posts that contain any of the
entries in our final list of n-grams. Unfortunately, a manual inspection
revealed several false positives --- \ie n-grams that were used outside the
context of feminist discourse. For example, the word `abortion' was found to
be used in the context of space travel missions and bigram instances of `me
too' were found to be used for simple agreement with statements unrelated to
feminism.

\para{{\it Reducing the false-positive rate.}}
In an effort to reduce the false-positive rates of our feminist discourse
dataset, we clustered all identified comments and posts by the context in which
they were used. This was done by using pre-trained BERT models
\cite{devlin2019bert} to convert each text submission into a vector form. These
vector forms automatically encode the context in which the contained n-gram was
used. Next, $k$-means clustering was used to cluster these vectors (and their
associated texts) into 100 clusters. $k=100$ was selected based on its
maximization of the silhouette coefficient metric \cite{clustermetric}. Next,
for each of the 100 clusters, the authors manually analyzed ten randomly
sampled texts and labeled them as on-topic or off-topic. All texts in the
clusters having only off-topic samples were then discarded as false-positives.

\para{Selection of Reddit users for our analysis.}
By parsing the meta-data of all on-topic clusters identified in the previous
step, we identified a total of 779K Reddit users who had engaged in some
discussions (\ie had at least one post or comment submission) related to
feminism between 01/2015 and 08/2020. 
Since our study requires us to achieve statistical significance in our analysis
of user trait changes related to feminism, we discard all users having less
than 20 submissions related to feminism during this 68 month period.
After this filtering, we were left with 12.7K Reddit users whose engagement
with the Manosphere were the subject of this study.

\subsection{Identifying Manosphere communities} \label{sec:methods:subreddits}
Our work seeks to understand how users engaged in feminist discourse (obtained
from the methods described in \Cref{sec:methods:users}) are influenced by
engagement with the Manosphere on Reddit. This requires us to identify Reddit
communities that belong in the Manosphere. 
For this, we use prior research by Ribeiro \etal
\cite{Manosphere-subreddits,ribero2021evolution} which identified 56
subreddits (comprising of 22M posts and 779K unique users) as Reddit's
Manosphere. 
These communities were identified by following references to related subreddits
identified on the subreddit Wiki pages of the most notable known Manosphere
communities on Reddit, including \subreddit{Incels}, \subreddit{TheRedPill},
\subreddit{MGTOW}, and \subreddit{Braincels}.
Our study focuses on the 68-month period between 01/2015 and 08/2020 during
which only 50 of these communities were active. In the context of our study,
these 50 communities serve as the Manosphere.

\subsection{Identifying relationships between events and traits}
\label{sec:methods:relationship}
\label{sec:methods:relationships}
In this section, we outline the methods used to identify relationships between
the events users experience and the traits they exhibit. In later sections, we
measure the events users experience monthly related to participation, interactions,
and perception. By identifying their relationship with each of the trait, we
answer our research questions seeking to identify the influence of social
engagements on warning behaviors. We employ two methods to determine whether an
event influences a trait: 1) Control-treatment comparative analysis and 2)
Cohort based regression analysis.

\para{Control-treatment comparative analysis.}
\label{sec:methods:comparative}
To determine whether an event influences a trait, we design a treatment-control
comparative analysis. All of our analyses are conducted on users active
throughout the time of our study (\ie from 01/2015 to 08/2020). The methods for
measuring traits and warning behaviors for each user in
\Cref{sec:background:traits} yields 8 time-series for each of the 8 traits
listed in \Cref{tab:background:traits}. Similarly, for each user, we have
collected the online events they experience along with the date of the event.
% The analysis
In our experiment, we seek to measure the effect of an event on the trait value
of a user. To this end, considering the event as the treatment, we compare the
post-event trait value of the treatment user with the post-event trait value of
a matched control user.
% How we match
We construct our treatment group from users who have experienced the event at
least once. Whereas, our control group consists of users who do not experience
the event at all. To isolate the influence of the event on the traits, for each
trait, we create treatment-control matched pair by matching each treatment user
with a control user based on the similarity of their pre-event trait values. We
define pre-event trait values as all monthly trait values since the first month
of our study (\ie 01/2015) until the month of the event.  To compute the
similarity between the pre-event trait values of treatment and control users, we
use Dynamic Time Warping (DTW) \cite{muller2007dynamic}.  DTW computes euclidean
distances between the points of two time-series and provides the total distance
between them as the sum of the minimum distances. To validate our matching,
we compute the average of the pre-event trait values for each of the user. Next,
we aggregate the pre-event trait averages of treatment and control users
separately. Using a test of significance, we determine whether the distribution
of pre-event trait averages are similar between the treatment and control
groups. Performing t-tests on these aggregates for each trait, we find all of
our traits have treatment and control groups with similar pre-event trait
averages.
% what is post trait
After constructing the matched treatment-control pairs for each of the trait,
we compare the post-event traits of the matched users to find any significant
differences. For each pair, we represent the post-event traits as the mean of
the trait values after the event date.
Finally, to measure the influence of the event, for each trait, we aggregate the
post-event trait from treatment users and compare them to the aggregated
post-event trait of the matched control users. For each trait, we compute the
average difference between these two aggregates and use t-tests to determine
whether the difference is significant. We report the average percentage
difference treatment user experience compared to the control user along with
whether the differences were statistically significant ($p> 0.05$).

\para{Cohort-based regression analysis}
\label{sec:methods:regression}
To establish any immediate influence between an event and the exhibited trait,
we design a cohort-based regression analysis. As mentioned in
\Cref{sec:methods:comparative} our method to measure trait values for each user
yields a time-series for each of the trait. Similarly, we measure the events
experienced by the users and represent them as time-series of events. In this
analysis, we seek to determine the average change in a trait value ($\Delta
t_i$) following an event in the prior month ($e_{i-1}$).
First, for each user, we compute the monthly change in their trait value
($\Delta t_i$) by performing the first order difference of the time series of
the trait. Next, for each user, we construct a set of tuples consisting their
event value for a month and the trait change the following month $(e_{i-1},
\Delta t_i)$. This yields a set of tuples for each trait for each user. Next,
for each of the trait, we aggregate the sets of tuples over all users giving us
a single set for each trait.
Given that for each trait, we have a set with data points containing the
magnitude of event experienced in a month and the change in the trait value
exhibited in the next month we seek model this relationship using a linear
We perform linear regression on the set of tuples with change in the trait
values ($\Delta t_i$) as the dependent variable and the event $e_{i-1}$ as the
explanatory variable ($\Delta t_i = \alpha + \beta \cdot e_{i-1}$). 
Coefficients of the event in the linear regression model $\beta$ represent the
influence of the event on the trait the following month. In our results we
report the coefficients as the percentage change in trait a single instance of
event influences along with whether the influence is statistically significant
($p<0.05$) as reported by the regression model.
Using linear regression to measure the influence of the
explanatory variable on the dependent variable has been well studied and
documented \cite{molnar2020interpretable}. Prior works have employed
regression to, for example, measure the influence social factors on joining a
community \cite{phadke2021makes}.

\section{Participation} \label{sec:participation}

\para{Overview.}
Online communities are fundamental to the structure of Reddit. Participation in
a community, by way of making a post or comment within it, signals an interest
in the discourse ongoing within it. 
In this section, we answer \textit{RQ1: How does participation in the
Manosphere communities influence the traits associated with Radicalization?} 
We analyze the influence of participation within the Manosphere at multiple
levels. Specifically, when measuring the influence of participation in the
Manosphere on a users' traits, we consider participation defined by
(\Cref{sec:participation:joining}) joining a Manosphere community, 
(\Cref{sec:participation:magnitude}) regular commenting and posting inside the
Manosphere, and (\Cref{sec:participation:epistemic}) the creation of epistemic
participation bubbles (\ie regular commenting and posting that is isolated to
the Manosphere).
We note that our definition does not include observational engagement in the
community where the user does not make a submission and only browses the
content (\ie `lurking'). 
For our analysis to uncover the influence of each type of participation
on warning behaviors described in \Cref{sec:background:traits}), we use
a combination of the control treatment analysis and cohort-based regression
analysis methods described in \Cref{sec:methods:relationships}.
A summary of all our results is provided in \Cref{tab:participation}.

\begin{table}[t]
\centering
\small
%\resizebox{\textwidth}{!}{%
\begin{tabular}{lcccc}
\toprule
  {Warning behavior}   & \multicolumn{1}{c}{{Joining (\Cref{sec:participation:joining})}}  
                       & \multicolumn{1}{c}{{Regular participation (\Cref{sec:participation:magnitude})}} 
                       & \multicolumn{2}{c}{{Epistemic bubbles (\Cref{sec:participation:epistemic})}} \\
  {}                   & $\Delta_{CT}$ & $\Delta_{CR}$  & $\Delta_{CT}$ & $\Delta_{CR}$ \\  
  \midrule
  Fixation    & {\bf 65}    & {\bf 0.80}  & {\bf 125} & {\bf 2.80}  \\ 
  Grievance   & {\bf 4}     & {\bf 0.05}  & {1}       & {0.02}      \\ 
  Power       & {\bf 6.9}   & {0.05}      & {1}       & {0.04}      \\
  \midrule
  Anger       & {\bf 19}    &	{0.16}      & {\bf 8}   &	{0.12}      \\
  \midrule
  Negativity  & {\bf 6.5}   & {\bf 0.02}  & {\bf 4}   & {\bf 0.03}  \\
  Toxicity    & {\bf 24}    & {\bf 0.03}  & {\bf 12}  & {\bf 0.11}  \\
  \midrule
  In-group    & {-3.4}      &	{-0.01}     &	{\bf -9}  &	{0.1}       \\
  Out-group   & {\bf 10.8}  &	{0.07}      & {\bf 7}   & {0.02}      \\
  \bottomrule
\end{tabular}
  \caption{Summary of results: Influence of participation in the Manosphere on
  warning behaviors. $\Delta_{CT}$ represents the measured \emph{percentage
  difference} in the corresponding trait between the treatment and control
  group in the control treatment analysis. $\Delta_{CR}$ represents the
  \emph{percentage change in the trait per unit increase in the participation
  metric} in our cohort-based regression analysis. {\bf Bold values} indicate
  statistically significant ($p$ < 0.05) differences.}
  \label{tab:participation}
%}
\end{table}

% The magnitude of their participation is represented by the number of submissions
% made in that community per month. An increase in the number of submissions would
% suggests a higher level of participation and therefore an increase in engagement
% in the topic of community. \note{add what this
% does and why this was done???.} To measure the participation in the Manosphere,
% for each month, we measure the number of submissions made by each user in
% \Fusers in any of the community in the Manosphere. This yields a single
% time-series for each user representing the magnitude of their participation in
% the Manosphere over time.
% 

\subsection{Influence of joining the Manosphere}
\label{sec:participation:joining}

We begin our study on the influence of participation in the Manosphere by
analyzing how the act of joining the Manosphere impacts warning behaviors of
users. 
We say that a user has joined the Manosphere when they make their first comment
or post within any Manosphere community.  

\para{Measuring the impact of joining the Manosphere.}
To measure the impact of joining the Manosphere on the warning behaviors, we
perform our comparative control treatment experiment described in
\Cref{sec:methods:relationship}. In this case, the `treatment' is considered to
be the event of joining the Manosphere.
We first create our treatment group by selecting all users who joined the
Manosphere between 06/2015 and 02/2020. Next, we match each treatment user with
a control user, who has never participated in the Manosphere, based on the
similarity of their history of warning behaviors prior to the joining date of
the treatment user.  
This user similarity is computed using Dynamic Time Warping (DTW)
\cite{muller2007dynamic} as described in \Cref{sec:methods:relationship}.
A KS-test confirms the goodness of this matching by finding no statistical
difference between the two groups, along any warning behavior, prior to the
treatment being applied.
Our user matching is a best-effort attempt at removing any latent confounding
variables that might impact our control treatment comparative analysis. 
Finally, for each of the warning behaviors, we measure the difference in the
values of the treatment and control group after the treatment's joining date and
determine whether the difference is statistically significant using a $t$-test
($p$ < 0.05). 
Observing a statistically significant change in any of the warning behaviors
would suggest influence from joining the Manosphere on that warning behavior.

\para{Result: Joining the Manosphere significantly increases the prominence of
nearly all warning behaviors.}
The results of our control treatment comparative analysis are illustrated in
the \emph{Joining} column in \Cref{tab:participation}. The results are alarming
and show that the act of joining the Manosphere, by way of making a post or
comment submission, results in the subsequent statistically significant
increase in all but one warning behavior (in-group identification).
Most prominently, in comparison to their similar counterparts in the control
group, we see that users who join the Manosphere exhibit a 65\% increase in
fixation on feminist discourse, 24\% increase in submissions classified as
toxic, and 19\% increase in usage of words associated with anger.
Further, the act of joining the Manosphere appears to result in the formation
of an `out-group' which increased by 10.9\%. 
These results support prior research by ethnographers that have suggested that
the Manosphere now serves as a platform devoted (increased fixation) to
expressing anger and hatred (increased anger and toxicity) towards women
(increased out-group identification). 

\subsection{Influence of regular participation in the Manosphere}
\label{sec:participation:magnitude}
Our previous results have shown that the act of joining the Manosphere results
in notable increases in warning behaviors --- particularly fixation, toxicity,
and anger. We now use our cohort-based regression analysis to understand how
the \emph{magnitude of participation} within the Manosphere influenced these
traits. Put another way, how much of an impact does a single Manosphere
participation event (in this case, a post or comment submission) have on
a users' warning behaviors?

\para{Measuring the impact of a single act of participation.} 
We now focus exclusively on the users who have participated in the Manosphere.
We use the regression analysis approach described in
\Cref{sec:methods:relationship}.
We consider the submission of each post or comment as an `participation event'.
Then, for each of the 68 months in our study and for each Manosphere
participant, we compute the number of participation events occurring in that
month. Finally, we use our month-to-month trait change measurements computed
for each of these users, to obtain a set of ($\delta_{\text{trait}}$,
$\text{event}_{\text{participation}}$) tuples where each tuple represents the
change in a trait from the prior month and the number of participation events
that occurred in the prior month.
We use these tuples to compute the regression co-efficients associated with
each trait using the trait changes as the dependent variable and the
participation events as the exploratory variable. 
The regression coefficient associated with each trait represents the percentage
increase that occurs in the corresponding trait from each participation event.
A statistically significant ($p$ < .05) positive coefficient for a warning
behavior would suggest that each submission in the Manosphere positively
influences the corresponding warning behavior.

\para{Results: Even a single participation event can increase language- and
outlook- based warning behaviors.}
The results from our regression analysis are illustrated in the \emph{Regular
participation} column in \Cref{tab:participation}. Traits in the language and
outlook warning behaviors alone observe any statistically significant increases
from a single participation event. Specifically, fixation (+0.8\%), negativity
(+0.02\%), toxicity (+0.03\%) and grievances (+0.05\%) all have statistically
significant changes per submission.
These increments can be explained as a consequence of adopting ideologies from
the Manosphere and conforming to their norms.
Further, they serve to highlight how delayed platform administration and
moderation decisions can harm the quality of online discourse and our ability
to prevent radicalization. After all, in some respects, these measured
coefficients show the harms caused to the user and the broader community that
occur from allowing a single participation event in the Manosphere.

\subsection{Influence of epistemic participation bubbles in the Manosphere}
\label{sec:participation:epistemic}

Prior work on radicalization and extremism have frequently demonstrated the
influence of epistemic bubbles in the adoption of extreme ideologies. 
Epistemic bubbles occur when users isolate themselves, either voluntarily or
through algorithmic personalization, by limiting the sources of information to
a selected few conforming to their beliefs. This often leads to the user
developing a distorted view of reality which facilitates radicalization. 
For our analysis, we determine the impact of disproportionately high
participation in the Manosphere on each of the warning behaviors. By
participating exclusively in the Manosphere, a user could develop a depraved
and dangerous understanding of feminism which could then transform into
misogynistic radicalization. 

\para{Measuring the impact of epistemic participation bubbles.}
To measure the influence of isolated and disproportionate participation in the
Manosphere, we first measure the proportion of each users participation inside
the Manosphere. A users proportion is represented by the percentage of their
total submissions made inside the Manosphere in a month. A higher percentage
would suggest lack of diversity in their community participation. 
We perform two analyses: First, we conduct a control treatment analysis, as
described in \Cref{sec:methods:relationships}, to identify how traits are
influenced in users with higher than median participation in the Manosphere.
Second, we perform a regression analysis, as described in
\Cref{sec:methods:relationships}, to uncover the changes in traits caused by
a single percentage increase in the proportion of Manosphere participation.

\para{{\it Control treatment analysis.}}
We consider the `treatment' in our analysis to be a high ratio of participation
in the Manosphere. Therefore, as our treatment group, we select the Manosphere
participants with higher than median participation ratios. Using the same
matching methodology as before, we construct our control group by matching each
of the treatment users with a Manosphere participant not belonging in the
treatment group and having the most similar measures of warning behaviors prior
to joining the Manosphere. We then compare the measures of the warning
behaviors occurring in the control and treatment group and test the
significance of their differences.

\para{{\it Regression analysis.}}
For this analysis, we consider an event to be a unit percentage increase in the
ratio of participation in the Manosphere. Then, using the same approach as
before, we compute the regression coefficient associated with each trait. This
coefficient represents the percentage change in the trait value per unit
percentage increase in a users' Manosphere participation ratio.

\para{Results: Disproportionate participation increases fixation and
outlook-based warning behaviors.}
The results from our analyses are illustrated in the \emph{Epistemic bubbles}
columns in \Cref{tab:participation}. Similar to our previous findings, the most
notable increases in both our analyses are in the feminism fixation trait. Our
control treatment analysis shows that users exhibiting higher than median
Manosphere participation ratios are 125\% more fixated on feminism than their
similar counterparts and a unit percentage increase in this ratio is associated
with a 2.80\% increase in the users' fixation. 
Similarly, users with higher than median ratios of Manosphere participation
also appear to exhibit more negativity (4\%) and toxicity (12\%) than their
counterparts. Further, each unit percentage increase in this ratio is
associated with a 0.03\% and 0.11\% increase in these traits, respectively.
These results highlight the importance of avoiding epistemological bubbles and
support prior research highlighting the role of such bubbles in polarization
and radicalization.

\subsection{Takeaways}
Taken all together, our results show that engaging, by way of participation,
with Manosphere communities does cause significant increases in traits
associated with radicalization. This finding suggests the role that these
communities play in contributing to our increasingly radical and polarized
discourse. Further, our results highlight the importance of effective and
timely platform moderation and administration --- after all, we see that even
small amounts of participation have a significant influence on a user's warning
behaviors.

\section{Interaction} \label{sec:interaction}

\para{Overview.} In \Cref{sec:participation} we exclusively focused on the
influence of user participation \emph{inside} a Manosphere community. In this
section, we focus on understanding how traits are influenced by repeated
interactions with \emph{elite} members of the Manosphere \emph{on communities
outside the Manosphere}. This allows us to answer \textit{RQ2: How does
repeated interaction with elite Manosphere members outside the Manosphere
influence the traits associated with radicalization?} 
This question is motivated by prior literature which, in the context of
conspiracy theories, demonstrated the important role of online user-to-user
interactions in influencing users to join communities or adopt ideologies
\cite{phadke2021makes}.
For our purposes, we consider comments made on another users' posts and replies
to other users' comments as an interaction. These interactions imply user
engagement with other individuals on a particular topic. 
We characterize each user-to-user interaction into one of two groups based on
the direction of interaction with the Manosphere member: 
(\Cref{sec:interaction:soliciting}) a solicitation interaction where
a non-member user initiates the interaction with a Manosphere member and
(\Cref{sec:interaction:recruiting}) a recruitment interaction where
a Manosphere member initiates the interaction with a non-member user.
Interactions can be initiated by replying to a post or comment
Since we are only interested in interactions involving elite Manosphere
members, we restrict our analysis to interactions involving the top 10\% of
most active Manosphere members who demonstrate any warning behavior in the top
10\% of all other Manosphere members. In total, this included 931K
interactions involving 1,325 Manosphere members and 5.3K other
users.
In our analysis, we use our control treatment comparative and cohort-based
regression analysis to understand the influence of each of these interactions.
A summary of our results is provided in \Cref{tab:interaction}.

\begin{table}[t]
\centering
\small
%\resizebox{\textwidth}{!}{%
\begin{tabular}{lcccc}
\toprule
  {Warning behavior}   & \multicolumn{2}{c}{{Solicitation interactions (\Cref{sec:interaction:soliciting})}}
                       & \multicolumn{2}{c}{{Recruitment interactions (\Cref{sec:interaction:recruiting})}} \\
  {}                   & $\Delta_{CT}$ & $\Delta_{CR}$  & $\Delta_{CT}$ & $\Delta_{CR}$ \\
  \midrule
  Fixation    & {\bf 150}   & {\bf 0.9}  & {\bf 99} & {\bf 0.13}  \\
  Grievance   & {\bf 4}     & {\bf 0.12} & {5}      & {\bf 0.05}      \\
  Power       & {\bf 8}     & {\bf 0.13} & {\bf 10} & {0.04}      \\
  \midrule
  Anger       & {\bf 16}    & {\bf 0.15} & {\bf 25} & {0.02}      \\
  \midrule
  Negativity  & {\bf 16}   & {\bf 0.04}  & {\bf 5}   & {\bf 0.01}  \\
  Toxicity    & {\bf 14}   & {\bf 0.05}  & {\bf 7}   & {0}  \\
  \midrule
  In-group    & {-3}      & {\bf 0.17}     & {\bf -2}  & {0.07}       \\
  Out-group   & {\bf 8}   & {\bf 0.07}     & {\bf 21}  & {0.02}      \\
  \bottomrule
\end{tabular}
  \caption{Summary of results: Influence of interaction in the Manosphere on
  warning behaviors. $\Delta_{CT}$ represents the measured \emph{percentage
  difference} in the corresponding trait between the treatment and control
  group in the control treatment analysis. $\Delta_{CR}$ represents the
  \emph{percentage change in the trait per unit increase in the interaction
  metric} in our cohort-based regression analysis. {\bf Bold values} indicate
  statistically significant ($p$ < 0.05) differences.}
  \label{tab:interaction}
%}
\end{table}

% 
% \parait{Identifying elite Manosphere members.}
% First, we define elite Manosphere members. As established earlier in ???, warning
% behaviors from the threat assessment toolkits are used to identify radicalized
% users. High warning behaviors exhibited by an individual online would be enough
% to flag the user as being potentially radicalized. Similarly, due to the
% misogynistic beliefs present at the core of the Manosphere, high activity in the
% Manosphere would suggest the adoption of such beliefs. Taken together, we
% define elite Manosphere members as users who are highly active in the Manosphere
% (more than the 90th percentile) and exhibit high warning behaviors (more than
% the 90th percentile).

\subsection{Influence of solicitation interactions}
\label{sec:interaction:soliciting}

A solicitation interaction is said to have occurred when a user initiates an
interaction with an elite Manosphere member. A key difference between
recruitment and solicitation, for our purposes, is that it is far more likely
for a user to have engaged with the elite Manosphere member on a particular
topic in the solicitation-type interaction as compared with recruitment. Our
analysis seeks to measure whether engaging with content submitted by elite
Manosphere members by interacting with it influences warning behaviors.

\para{Measuring the influence of soliciting interactions.}
We measure solicitation-type interactions by identifying and counting all
comments and replies made by users who have engaged in feminist discourse
(\emph{Cf.} \Cref{sec:methods:users}) to a submission made by an elite
Manosphere user in a community outside the Manosphere. We then count each of
these interactions as an event and compute the number of such events in each of
the 68 months of our study. The median number of solicitation interactions that
users in our dataset have with elite Manosphere users is 3/month.

\parait{Control treatment analysis.}
We consider the `treatment' in our analysis to be a higher-than-median number
of solicitation interactions with elite members of the Manosphere in
communities outside the Manosphere. 
Therefore, our treatment group consists of users who have more than three
solicitation-type interactions per month with elite Manosphere members. The
time at which the `treatment' is applied for these users is the date on which
their third solicitation interaction occurs.
For each of these treatment group users, we use our standard matching process
to identify the user who has had \emph{no interaction} with elite Manosphere
members and has the most similar measures of warning behaviors (in the months
prior to the treatment being applied to the treatment user). We add these users
to our control group. 
We find that the pre-treatment distributions of each warning behavior are
statistically indistinguishable using the KS-test for goodness of fit ($p$
< .05).
Next, we compare the distributions of each warning behavior between the
treatment and control groups to find any significant differences that emerge
after the application of the treatment.
Observing a statistically significant post-treatment difference in any of the
measured warning behaviors between the two groups would suggest an influence
from multiple solicitation-type interactions with Manosphere elites.

\parait{Regression Analysis.}
For our regression analysis, we consider an event to be a solicitation-type
interaction with a Manosphere elite. 
For each user in our dataset having a solicitation-type interaction with
a Manosphere elite outside the Manosphere, we count the number of these events
occurring during each month. 
We then use the trait measure changes that occur within each month to build the
$(e_{i-1}, \Delta t_i)$ tuples which represent the change in a trait from the
prior month and the number of solicitation-type interaction events that occurred
in the prior month for each user and for each month in our dataset.
Finally, we compute the linear regression coefficient for each trait. This
coefficient represents the change in trait value observed per unit increase in
number of solicitation-type interactions with Manosphere elites. 

\para{Results: All warning behaviors are exacerbated by solicitation
interactions with Manosphere elites.}
The results from our analyses are illustrated in the \emph{Solicitation
interactions} columns in \Cref{tab:interaction}. 
We find that all traits are positively influenced by solicitation-type
interactions with elite Manosphere members that occur outside the Manosphere. 
The most notable increases occur in the traits associated with fixation, anger,
and the outlook-related traits of negativity and toxicity --- with our
treatment users showing nearly $2.5\times$ higher fixation and 14-16\% higher
anger, negativity, and toxicity than users who never interacted with Manosphere
elites.
Our regression analysis shows that even the cost of a single solicitation-type
interaction with Manosphere users is high and causes statistically significant
increases of 0.05\% - 0.90\% in warning behavior measures.
These results are similar to our findings on the influence of joining
a Manosphere community. What is more concerning here, however, is the fact that
even interactions that: (1) are not initiated by the Manosphere elite and (2)
occur outside the Manosphere can have such a pronounced influence on a user.
In fact, further analysis shows that a large number of these solicitation
interactions occur in general subreddits such as \subreddit{AskReddit},
\subreddit{politics}, and \subreddit{unpopularopinion} where 4.5\%, 3.3\%, and
1.3\% of all solicitation interactions took place. However, the large majority
of solicitation interactions were generally found to be in other known toxic
communities such as \subreddit{KotakuInAction} which was the original home of
GamerGate, \subreddit{TumblrInAction}, and \subreddit{The\_Donald}.

\subsection{Influence of recruitment interactions}
\label{sec:interaction:recruiting}
A recruitment interaction is an interaction which is initiated by the elite
Manosphere member. Prior research focusing on the radicalization of incels and
white supremacists note that coordinated individuals in extremist online
communities are common to aid the spread of their ideologies
\cite{wong2015supremacy,evans2018memes}. 
Such efforts involve the recruiters' participation in mainstream (often,
non-political) communities where they share their grievances and ideologies in
attempts to influence users by way of eliciting sympathetic or agreement
reactions. 
As noted by accounts from ex-radicals, these recruitment attempts can often
lead to realization of false grievances and the subsequent adoption of dangerous
or extreme ideologies. 

\para{Measuring the influence of recruitment interactions.} 
Our approach for measuring the influence of recruitment interactions is similar
to the methods used for solicitation interactions. 
We measure recruitment-type interactions by identifying and counting all
comments and replies made by Manosphere elites to posts or comments from users
who have engaged in feminist discourse (\emph{Cf.} \Cref{sec:methods:users}) in
communities outside the Manosphere. We consider each of these interactions as
an event and compute the number of such events in each of the 68 months of our
study. The median number of recruitment interactions that users in our dataset
experience is 1/month.

\parait{Control treatment analysis.}
Similar to before, we consider the `treatment' in our analysis to be
a higher-than-median number of recruitment interactions with elite members of
the Manosphere. Therefore, our treatment group consists of all users who
experienced at least one recruitment interaction in a non-Manosphere community.
We then repeat the identical matching process to create a control group of
users with similar traits and no recruitment interactions. As before, we
measure the statistical significance of the differences in trait distributions
between the control and treatment groups prior to and after the treatment is
applied.
We then attribute any measured statistically significant difference between the
two groups after the treatment is applied to the treatment user being the
subject of a recruitment interaction with the Manosphere elite.

\parait{Regression analysis.}
We use the same approach as the regression analysis described for solicitation
interactions (\emph{Cf.} \Cref{sec:interaction:soliciting}) to obtain
$(e_{i-1}, \Delta t_i)$ tuples for each trait, user, and month ($i$).
Each of these tuples indicates a users' change in a measured trait value from the
prior month and the number of recruitment interactions experienced by the user
in the prior month.
As before, we compute the regression coefficient for these tuples to identify
the influence of a unit increase in the number of recruitment interactions on
each trait.

\para{Results: Recruitment interactions are generally less influential than
solicitation interaction but have a stronger influence on out-group
identification and anger.}
The results of our analyses are illustrated in the \emph{Recruitment
interactions} columns in \Cref{tab:interaction}.
We find that recruitment interactions also positively influence nearly all
warning behaviors. However, when comparing such interactions with solicited
interactions, we note several key differences. 
First, our regression analyses show that the influence of a single recruitment
interaction on a user is much lower than the influence of a single solicitation
interaction, while our treatment control analysis results are generally in the
same ballpark. This suggests that recruitment efforts might be targeted and
require multiple contacts with the user.
Second, our control treatment analysis shows that users in our treatment group
experience much higher increases in the measures of anger and out-group
identification when they experience a recruitment interaction (compared to
a solicitation interaction). This suggests the nature in which recruitment
interactions are effective --- they are able to cause their subjects to exhibit
anger and instigate the formation of an out-group. Although, based on our
analysis we cannot conclude whether their texts achieve both simultaneously and
whether the target of their anger and constructed out-group are women.
Supporting previous theories on recruitment into radical groups, further
analysis shows that the majority of such interactions occured in general
subreddits such as \subreddit{AskReddit}, \subreddit{politics},
\subreddit{news}, \subreddit{worldnews}, \subreddit{movies}, \subreddit{pics},
and \subreddit{nfl}. These mainstream subreddits together were the home to 16\%
of all identified recruitment interactions.

\subsection{Takeaways}

Our study on the influence of interaction with Manosphere elites shows that
they are very effective at increasing measures of warning behaviors. We also
uncover key differences in the effects of interactions that are initiated by
the user (solicitations) and interactions that are initiated by the elite
(recruitment). Notable amongst these is the venue in which interactions occur
and the warning behaviors that they exacerbate. We hypothesize that some of
these differences can be attributed to the `control' that elites possess over
recruitment interactions which allows them to target vulnerable individuals or
ideological sympathizers.

\section{Perception}
\label{sec:perception}

\para{Overview.}
Social status is a critical component in Reddit's structure. Being
a crowdsourced content aggregator, Reddit relies largely on the concept of
social status to motivate users to create and share content. The social status
system on Reddit is composed of \textit{upvotes}, \textit{downvotes}, and
\textit{Karma} which is the sum of upvotes and downvotes received by the user
across all of their submissions. 
Other than determining the visibility of the content, the score a submission
receives also reflects the value it provides for the community. 
Apart from motivating users to participate social status systems can also be
exploited or can cause adverse effects on user behavior. Social science and
psychology research studies have shown the significant role of social rejection
in the process of radicalization and adoption of extreme ideologies
\cite{renstrom2020exploring, haddad2021online, jasko2017quest}. 
Similarly, social acceptance by extremist groups might also have radicalizing
influence. 
In this section, motivated by the above theories and using Reddit's voting
mechanism and karma as a proxy for social status, we focus on answering
\textit{RQ3: How does social status within and outside the Manosphere influence
traits associated with radicalization?}
In our analysis, we consider the influence of:
(\Cref{sec:perception:rejection}) rejection from non-Manosphere communities
which we measure by counting the number of user submissions having negative
karma --- \ie more downvotes than upvotes which results in it being hidden from
other Redditors and (\Cref{sec:perception:acceptance}) acceptance from
Manosphere communities which we measure by counting the number of user
submissions having higher-than-median karma within the Manosphere.
Similar to our methods in \Cref{sec:interaction}, we conduct a control
treatment comparative and cohort-based regression analysis. A summary of our
results is provided in \Cref{tab:perception}.

\begin{table}[t]
\centering
\small
%\resizebox{\textwidth}{!}{%
\begin{tabular}{lcccc}
\toprule
  {Warning behavior}   
  & \multicolumn{2}{c}{{Rejection outside the Manospehere(\Cref{sec:perception:rejection})}} 
  & \multicolumn{2}{c}{{Acceptance in the Manosphere (\Cref{sec:perception:acceptance})}}\\
  & $\Delta_{CT}$ & $\Delta_{CR}$  & $\Delta_{CT}$ & $\Delta_{CR}$ \\
  \midrule
  Fixation      & {\bf 35} & {\bf 0.6}  & {\bf 137}   & {\bf 1.44}\\
  Grievance     & {\bf 2}  & {  -0.01}  & {\bf 6}     & {\bf 1.4} \\
  Power         & {\bf 3}  & {0.03}     & {\bf 8}     & {\bf 1.3} \\
  \midrule                                                       
  Anger         & {\bf 16} & {0.13}         & {\bf 18}    & {\bf 2.1} \\
  \midrule                                                       
  Negativity    & {\bf 14}   & {\bf 0.19} & {\bf 8.5}   & {\bf 0.66}  \\
  Toxicity      & {\bf 31}   & {\bf 0.25} & {\bf 25}    & {\bf 1.25}  \\
  \midrule                                                       
  In-group      & {\bf -7}  & {0.01}   & {3}         & {-0.73}       \\
  Out-group     & {\bf 3}  & {0.03}    & {\bf 15.6}  & {\bf 1.59}  \\
  \bottomrule
\end{tabular}
  \caption{Summary of results: Influence of social status within and outside
  the Manosphere on warning behaviors. $\Delta_{CT}$ represents the measured
  \emph{percentage   difference} in the corresponding trait between the
  treatment and control   group in the control treatment analysis.
  $\Delta_{CR}$ represents the   \emph{percentage change in the trait per unit
  increase in the perception metric} in our cohort-based regression
  analysis. {\bf Bold values} indicate   statistically significant ($p$ < 0.05)
  differences.}
  \label{tab:perception}
%}
\end{table}

\subsection{Influence of social rejection from non-Manosphere communities}
\label{sec:perception:rejection}
In this section we seek to determine whether being negatively received by the
general public can facilitate increases in the warning behaviors associated
with radicalization through feelings of rejection and
alienation. As mentioned earlier, research into the process of radicalization
has noted the role of social rejection as a pathway to radicalization. In our
analysis, we measure whether being negatively received from outside of the
Manosphere can influence a user's warning behaviors. 

\para{Measuring the influence of social rejection.}
We use Reddit's content voting mechanics to obtain a proxy for social
rejection. We begin by analyzing the number of upvotes and downvotes received
in each post or comment submission made by all users who participated in
feminist discourse (\textit{Cf.} \Cref{sec:methods:users}). 
For each of these users, we count the number of submissions made outside the
Manosphere that received a greater number of downvotes than upvotes. We
consider each of these submissions to constitute one event of social rejection.
We make this characterization because: (1) such an event signifies general
mainstream disagreement with the content of the submission and (2) submissions
that have a negative score are generally hidden from view by Reddit.
We count the number of such events that occur for each month in our study for
each user in our dataset. A higher number of such events occurring in a month
signifies a higher experience of social rejection from non-Manosphere
communities.
We find that the median number of rejection events experienced by users in our
dataset was 5/month.
Note that in this analysis, we do not distinguish between members of the
Manosphere and other users. Instead, we are interested in obtaining a general
understanding of how social rejection influences warning behaviors.

\parait{Control treatment analysis.}
We consider the `treatment' in our analysis to be a median or
higher-than-median number of social rejection events occurring outside the
Manosphere communities.
Therefore, our treatment group consists of all users who experienced at least
five rejection events per month outside the Manosphere. The time at which the
treatment is applied is the date on which their fifth rejection event occurs.
For each of these treatment users, we use our standard matching method to find
a user who has less than five social rejection events and shares similar
measures of warning behaviors as the pre-treatment treatment user. We add this
user to our control group.
We then use the KS-test to verify the goodness of fit of the pre-treatment
distributions of the control and treatment group ($p$ < .05). Following this,
we compare the distributions of each warning behavior measured post-treatment
for the two groups. Observing statistically significant differences here would
suggest an influence of social rejection on the corresponding warning
behaviors.

\parait{Regression analysis.}
We use our standard regression analysis set up to derive
($\delta_{\text{trait}}$, $\text{event}_{\text{rejection}}$) tuples for each
trait, user, and month. Each of these tuples indicate a users' change in
a measured trait value from the prior month and the number of social rejection
events experienced by them in the prior month. Finally, we compute the linear
regression coefficient for these tuples to identify the influence of a unit
increase in social rejection events on each warning behavior.

\para{Results: Social rejection strongly influences outlook-based warning
behaviors.}
The results of our analyses are illustrated in the \emph{Rejection outside the
Manosphere} columns in \Cref{tab:perception}. Once again, in our control
treatment analysis, we find that social rejection has a significant influence
on all measured warning behavior. Our regression analysis, on the other hand,
shows that the changes in warning behaviors for a single rejection event are
marginal and statistically insignificant except in the cases of fixation,
negativity, and toxicity.
Notable among these influences is the sharp increase in the outlook-based
warning behaviors of negativity and toxicity. The increases in these traits are
much higher for social rejection events than any of the other events considered
in our research. The control treatment analysis shows that users who experience
higher-than-median social rejection in the prior month will demonstrate a 31\%
higher measure of toxicity in the following month when compared to a similar
user in the control group. Along similar lines, we see that each recorded
rejection event experienced by a user causes a 0.25\% increase in their
toxicity measure for the following month.
Interestingly, we find that rejection events result in a decrease in measured
in-group identification (7\% lower than control group), suggesting that
rejection in fact might result in loss of a sense of belonging. 
As a whole, these results suggest that platforms that provide mechanisms for
social rejection of users might only spur the rejected user to become
increasingly toxic, fixated, angry, and negative.

\subsection{Influence of social acceptance from the Manosphere}
\label{sec:perception:acceptance}
Research has shown that individuals seeking acceptance, validation for their
grievances, and a sense of belonging will change their own behaviors and adopt
ideologies of a group that demonstrates social acceptance. 
We test this insight in the context of the Manosphere by analyzing if being
positively received by the Manosphere can facilitate increases in the warning
behaviors associated with radicalization.

\para{Measuring the influence of social acceptance.}
To measure social acceptance inside the Manosphere, we collect all submissions
made in the Manosphere for each month. Next, we calculate the median karma
score received by each comment or post submission. This score was 28.75 votes
--- \ie, the median difference between the number of upvotes and downvotes
received by a submission was 28.75. For each Manosphere member,
we then count the number of submissions that received a higher-than-median
karma score. We define each of these as a social acceptance event. We make this
characterization because: (1) such an event signifies agreement with the
community at large and (2) posts and comments with higher karma scores are
promoted in subreddits and threads, making them more visible to more users.
We find the median number of acceptance events experienced by Manosphere
members inside the Manosphere is 1/month.

\parait{Control treatment analysis.} 
We consider the `treatment' in our analysis to be a median or
higher-than-median number of acceptance events occurring inside the Manosphere.
Therefore, our treatment group consists of all users who experienced at least
one acceptance event inside the Manosphere. The time at which the treatment is
applied is when the first acceptance event has occurred. 
Next, we match each treatment user with another Manosphere member who shares
similar pre-treatment warning behavior measures but has no acceptance events
inside the Manosphere. These matched users are added to our control group.
As before, we validate our groups using a goodness of fit test on their
pre-treatment behaviors and compute the group differences and statistical
significance in their post-treatment warning behaviors. 
Observing statistically significant differences here would suggest an influence
of social acceptance in the Manosphere on the corresponding warning behaviors.

\parait{Regression analysis.}
We use our standard approach to derive the ($\delta_{\text{trait}}$,
$\text{event}_{\text{acceptance}}$) tuples for each trait, user, and month.
Each of these tuples indicates a users' change in a measured trait value from
the prior month and the number of social acceptance events experienced by them
in the prior month. Finally, we compute the linear regression for this
collection of tuples to identify the influence of a unit increase in social
acceptance events on each warning behavior.

\para{Result: Social acceptance in the Manosphere can lead to high increases in
measures of warning behaviors even with a few occurrences.}
The results of our analyses are illustrated in the \emph{Acceptance in the
Manosphere} columns of \Cref{tab:perception}.
As expected, we see that fixation is most influenced by social acceptance in
the Manosphere with our treatment users experiencing a nearly 2.4$\times$
higher post-treatment measure in comparison to the control users. Each social
acceptance event corresponds to a 1.4\% increase in fixation on feminism. Other
highly influenced traits are toxicity (25\% more than control group and 1.25\%
increase per event), anger (18\% more than control group and 2.1\% increase per
event), and out-group identification (15.6\% more than control group and 1.59\%
increase per event). 
Of note is the influence of a single acceptance event on the traits exhibited
in the following month. 
Our regression analysis shows that a single acceptance event can have orders
of magnitude higher influence on warning behaviors than a single instance of
any other event considered in our study.
Generally, our results support previous research indicating that group
acceptance can lead to the adoption of the group ideology and behavior.

\subsection{Takeaways}
Our analysis shows that community perception of a user can be a significant
driver of user behavior. Specifically, we find that Manosphere users who
experience high amounts of rejection are likely to exhibit high increases in
negativity, toxicity, and anger.  
Unfortunately, we also see that just a single
instance of social acceptance inside the Manosphere can cause significant
increases in all warning behaviors --- suggesting the effects of validation on
extremist behavior can result in its stronger exhibition.
Taken together, these results suggest the importance of community behavior in
preventing radicalization. 

\section{Related Work}
Our work serves to better understand how online platforms facilitate the
development of traits and warning signs associated with radical and extremist
behavior. More specifically, we make two key contributions in the research area
of understanding the adoption of extreme ideologies. First, our work explores
the role of social engagements fundamental to online platforms as
pathways towards adoption of radical and extremist behavior. Second, our work
contributes to studying online misogyny and anti-feminism by exploring how
social engagements around the Manosphere can facilitate misogynistic
radicalization.

%\para{Threat Assessment Toolkits.}
%Research in the field of radicalization has been generally focused on the goal
%of developing theories and tools to identify radical behavior. Backed by The
%Patriot Act the motivation to preemptively identify terrorists and the access to
%vast information accelerated the research into radicalization.
%More recently, motivated to identify potential domestic threats, researchers
%have designed tools to identify radical and extremist behavior by monitoring for
%identified warning signs. Meloy \etal developed a typology of warning behaviors
%identied by prior researchers through interviews and observations of identified
%threats and proposed the use of warning behaviors for threat assessment
%\cite{melyowarning}. Subsequently, the Terrorist Radicalization Assessment
%Protocol-18 (TRAP-18) was designed to be used as a threat assessment toolkit
%\cite{meloy2018operational}. Similarly, researchers used other sources and
%methods to develop threat assessment toolkits for the identification of
%extremist behavior through warning signs. These toolkits include the VERA
%2\cite{pressman2012calibrating}, ERG 22+ \cite{lloyd2015development} and IVP
%\cite{egan2016can}.

\para{Identifying Online Radicalization.}
Research on online radicalization has been generally focused on identifying radical
and radicalizing individuals online. It borrows heavily from prior research in threat assessment toolkits by
employing already validated warning signs curated in the tooolkits to identify
online radicalization. Borrowing from Meloy \etal, who motivated shifting the
use of threat assessment toolkits from a controlled environments to a more
normal and naturalistic environment, Cohen \etal have mapped a few relevant
warning signs categorized in TRAP-18 for the identification of extreme behavior
online \cite{cohen2014detecting}. The authors identify linguistic markers
present on social media for measuring Fixation, Identification, and Leakage. By
presenting these mappings, the authors enabled identification of potential
threats online. Furthermore, these mappings of warning signs for online
identification has been validated by the work done by Grover \etal. The authors
manually validate the association between linguistic markers and their relevant
warning behaviors after demonstrating presence of warning signs in the extremist
alt-right community on Reddit \cite{grover2019detecting}. Adding to the
linguistic markers for identification of extremist behavior online, van der Vegt
and colleagues introduced a psycho-linguistic dictionary to measure grievance
scores from submissions made on social media platforms \cite{van2021grievance}.
The authors evaluated the dictionary showing high performance in distinguishing
between lone-actor terrorist texts and neutral texts. Alternatively to the
threat assessment toolkits, researchers have also developed tools for
identification of radical users online that rely on participatory and textual
analysis. Scrivens \etal has developed a sentiment based identifier of radical
authors (SIRA). The authors use text from identified radical users to validate
and evaluate their identifier \cite{scrivens2015sentiment}. Subsequently, in
their later works, Scrivens \etal employ their sentiment based identifier to
identify and explore posting behaviors of right wing extremists
\cite{scrivens2021examining} where they identified large proportion of out-group
construction and attacks in the community. Our work borrows from the research
done to implement threat assessment toolkits and their warning signs as
linguistic markers to identify radical and extreme behavior online. The use of
these linguistic markers on Reddit and similar platform by prior works validates
their role as warning signs.

\para{Identifying Pathways of Radicalization} 
As algorithms role and importance in online ecosystem have increased,
research on auditing these black-boxes has been focusing on their
effects on user behavior, more specifically their ability to influence beliefs
and actions. Ribeiro \etal study the YouTube ecosystem for identification of
radical pathways towards Far-right content. Their analysis of user migration,
find a pipeline potentially aided by the recommendation algorithm of YouTube
between communities of increasing extremity \cite{ribeiro2020auditing}. A more
recent study by Munder \etal, however, failed to find evidence of a radicalizing
pipeline and argued the recommendation algorithm operates on a supply and demand
principle. Similarly, extensive research into the existence of echo-chambers and
filter bubbles show the impact of algorithmic personalization on an individual's
beliefs and actions \cite{nguyen2020echo, chitra2020analyzing}. Our work, on the
other hand, focuses examining the role of user-to-user social engagements in the
adoption of extreme ideologies.

\para{The Manosphere}
Research on exploring online anti-feminist movements and misogyny has focused on
understanding their behavior, dynamics and philosophy. Research
specifically into the Manosphere generally seeks to understand the ideologies
and behaviors of the communities inside. In their observational analyses of the
Manosphere, Farrell \etal show increasing hostility and violence towards women
online. Driven by lexical analysis, their work demonstrate the Manosphere use of
`flipping the narrative' technique and playing the victim in online communities
to legitimize their hatred towards women \cite{farrell2019exploring,
farrell2020use}. These findings support our results in
\Cref{sec:participation:joining} where we observe significant increase in
out-group construction, toxicity and anger after participation in the Manosphere
communities.
Ribeiro \etal perform a longitudinal study to reconstruct the history of the
Manosphere and observe its evolution over the years. Their work highlights the
growing toxicity and misogyny in Manosphere communities and the migration
patterns of users across communities within the Manosphere
\cite{ribero2021evolution}. In a subsequent work, Mamié \etal study the overlap
and migration between the anti-feminist communities and the far right. Their
cross-platform analysis on YouTube and Reddit demonstrate significant overlap
and find anti-feminist community to act as potential gateways to the Alt-right.
Massanari, in her work exploring GamerGate and The Fappening as two seminal
anti-feminist cases, notes the role of Reddit and its structure in
facilitating online misogyny and toxic cultures \cite{massanari2017gamergate}.
By studying the two cases, Massanari outlines how Reddit's design and governance
influenced the development, growth, and spread of `toxic-technocultures'.
Findings from our work follow these observations and seek to measure how the
social structures and systems fundamental to Reddit are exploited to spread
misogynistic ideologies. Supporting Massanari's claim we find Reddit's Karma
system and governance (or lack thereof) facilitating the spread of extreme
behaviors and ideologies.

\section{Discussion and Conclusions} \label{sec:discussion}

In this section, we highlight the limitations of our study and then place our
findings in the context of the future of platform governance and our
understanding of radicalization in the Manosphere.

\subsection{Limitations}
Our work is fundamentally a `best-effort' observational study aimed at
understanding how social user-to-user engagement with the Manosphere influences
an individuals' warning behaviors. Some of these limitations arise simply
because of the observational and textual nature of the data that we engage
with and others due to possibly incomplete datasets. 

\para{{Reliance on observational data.}}
The fact that our data is observational prevents us from performing ``gold
standard'' causal analysis experiments (\eg randomized control studies) that
allow us to draw indisputable and reliable conclusions regarding the effects of
engagement with the Manosphere. We navigate around this limitation by instead
using two well established statistical approaches (case control and
cohort-based regression analysis) to draw causal inferences from observational
data. Despite their wide use in prior observational data analysis and our best
efforts to consider latent confounders, these analyses approaches run the risk
of containing confounding biases that impact the conclusions of our study. 

\para{{Reliance on text analysis approaches to identify warning behaviors
and traits.}}
Identifying warning behaviors in an individual is a complicated process. The
warning behaviors used in our study were explicitly designed by psychologists
to be identified through mixed-method studies including surveys of and
interviews with the subject. 
Unfortunately, this is not feasible when seeking to understand the mechanics of
online radicalization at scale. In recognition of the importance of this
problem, previous research has validated many text analysis approaches for
measuring these warning behaviors. 
In our work, we restrict ourselves to only using these validated approaches and
the corresponding nine traits. Further, our goal is only to identify
engagements which result in a statistically significant increase in a measured
trait and \emph{not} to identify radicalized individuals. This allows our
trait measurement approaches some room for error since we only need them to be
reliable enough so that a \emph{statistically significant increase} in their
value maps to an increase in the users' real world exhibition of that trait.
Unfortunately, by only relying on validated approaches we are faced with an
incomplete picture of how the Manosphere influences individuals. For example,
we are unable to identify the influence of real world social factors. 

\para{{Possible incompleteness of our datasets.}}
Our analyses required us to perform: (1) filtering of users engaged in feminist
discourse and (2) identification of Manosphere communities on Reddit. Both are
possible sources of incompleteness in our analysis. 

\para{{\it Filtering users engaged in feminist discourse.}}
Our approach for identifying users engaged in feminist discourse required the
development of a seed list of n-grams from which snowballing was used to
develop a more complete list (\Cref{sec:methods:users}).
Despite our best efforts (which included validation from multiple
authors) to build a comprehensive seed list, it is possible that the set of
n-grams used in our filtering ignored hot-button topics in feminist discourse.
We note, however, that our goal in this filtering was to find users who were
engaged in feminist discourse and \emph{not to analyze the filtered comment and
post submissions}. Therefore, the impact of this source of incompleteness is
reduced by the high likelihood that our identified users may have engaged with
the topics not covered by our keyword list. This would only lead us to have
not included users who were solely engaged in a niche sub-topic related to
feminism.

\para{{\it Identifying Manosphere communities.}}
Our study seeks to understand how engagement with the Manosphere influenced
a users' warning behaviors. This required us to develop a set of communities
that belonged to the Manosphere. Our approach was to rely on a list of
subreddits curated by prior work \cite{ribero2021evolution,
Manosphere-subreddits}. Taken together, these communities involve
participation from 835K unique users. Although it is possible that this list is
incomplete, we believe that the list is sufficient to understand how engagement
with the Manosphere generally impacts user traits.

\subsection{Summary of results and their implications}
Despite our limitations, our research provides the following insights into how
user engagement with the Manosphere can influence warning behaviors associated
with radicalization.

\para{Participation in the Manosphere causes a significant increase in
language- and outlook-based warning behaviors (\Cref{sec:participation}).}
Our analysis shows how different dimensions of participation can influence
traits associated with radicalization. 
We see that the simple act of joining a Manosphere community can result in
significantly increased exhibitions of fixation on feminism, anger, toxicity,
and conceptualization of an out-group. 
Using our regression analysis, we show that after a user joins the Manosphere,
their continued participation in the Manosphere only exacerbates their radical
traits --- particularly fixation and toxicity.
Finally, as a user enters an epistemic bubble by participating
disproportionately in the Manosphere, the trend continues.

\parait{Implications.}
Our findings on the influence of participation highlight the important role
that platform administrators and moderators play in preventing the spread of
harmful ideologies and radical behavior.
Specifically, we see the potential impact that delayed community moderation
decisions can have on the users of a platform. After all, each participation
event that a user is able to have with a problematic community (in our work,
the Manosphere) appears to have a significant and harmful influence on multiple
aspects of their behaviors.
Unfortunately, the sheer size of online platforms results in the fundamental
impossibility of simultaneously achieving timely, fair, and high-quality
community moderation decisions.
Consequently, community moderation today remains largely reactive in the sense
that communities are allowed to exist until particularly egregious violations
are widely reported.
Recent research has shown, however, the promise of continuous tracking of
communities and their characteristics to help in the early identification of
dangerous communities. However, in the absence of nuanced intervention
approaches (\ie more fine-grained than banning or quarantining a community),
the applicability of these early identification approaches is limited.
Another approach that platforms might consider to mitigate the possibility of
a user engaging with harmful communities is via changes in platform mechanics
and exposure that users have to non-mainstream communities that they are not
a part of. 
As an example of one such success: until recently, the front page of Reddit
displayed a collection of posts from the entire platform based on their rising
popularity. This inadvertently became the subject of manipulation by
ideologically radical communities by having all their users simultaneously
engage with a post to have it reach the top of Reddit's front page which
subsequently led to large numbers of new users joining their community. This
problem was finally addressed by Reddit in 2020 when they changed the default
mechanics of the front page so that it only displayed content from white listed
subreddits. Consequently, the rate of users joining problematic communities
they previously were not aware of has reduced.

\para{Manosphere elites are effective propagators of radical traits even
outside the Manosphere (\Cref{sec:interaction}).} 
Our analysis on the solicitation and recruitment interactions with Manosphere
elites shows that they have strong influences on the warning behaviors of the
users they interact with, even when the interactions occur outside the
Manosphere. 
Specifically, we find that each solicitation interaction is associated with the
increase of all warning behaviors considered in our study --- most notable of
which is the sharp rise in fixation with feminism. 
Our analysis of the recruitment interactions shed light on the operations that
might be used to recruit vulnerable individuals into Manosphere communities.
For example, we see that the effects of such interactions are different than
the solicitation interactions --- with stronger effects in the anger and
out-group identification traits and lower per-interaction effect. This suggests
messaging strategies aimed at directing anger to a conceptualized group over
multiple contacts.
Further, we also see that the locations of such recruitment interactions are
general communities, whereas solicitation interactions occur in a mix of
general and other toxic communities.

\parait{Implications.}
Our findings regarding the power of Manosphere elites is particularly
concerning because of the fact that this power is effectively addressed even
outside the Manosphere. 
Beyond shedding light on how recruitment into extremist groups may occur, our
findings also bring to light the challenges associated with the (pseudo-)
anonymity provided by online platforms. 
Online platforms provide a safe harbor for interactions between individuals
whose motives and background remain largely unknown to each other. By
providing a short, effective, and accessible synopsis about a user's
participation history to any other user who might interact with them, it is
possible that the influence that extremist users possess will diminish.
Although Reddit does provide the ability to study the post and comment history
of an individual, this information is unavailable to users in a usable manner.
Instead, users have to click on a user profile link which navigates them away
from the page they are currently on and scroll through pages of posts and
comments to understand the background of the person they are interacting with.

\para{Warning behaviors are influenced heavily by community perception
(\Cref{sec:perception}).}
We find that users' warning behaviors are strongly influenced by whether they
are accepted or rejected by the community they participate in. 
Our analysis shows that rejection from communities outside the Manosphere
result in notably high expressions of negative outlook, toxicity, and anger
while simultaneously reducing in-group identification. This suggests that
strongly negative community reactions have a harmful effect that increase
emotional outbursts, unhealthy outlook, and isolation.
On the other hand, our analysis of social acceptance in Manosphere communities
where members already exhibit higher magnitudes of warning behaviors results in
a sharp rise in these behaviors --- particularly fixation, toxicity, and
out-group identification.

\parait{Implications.}
These findings suggest the need for improvements in fundamental mechanics of
online social platforms and their content curation systems.
Content curation systems in discussion-based platforms such as Reddit have
a heavy reliance on user-driven mechanism such as voting, commenting, or
liking. 
Although these mechanisms do facilitate more enjoyable experiences for the
majority of users of the platform, they present a problem. As our research
shows, they become problematic when they can be leveraged to explicitly signal
rejection from the community and subsequently result in increasingly
unhealthy reactions such as increasing the rejected user's warning behaviors. 
As demonstrated in our social acceptance analysis, they also pose a problem in
communities where extremist ideologies are the norm and questioning and
dissenting voices are suppressed (or literally hidden from the view of other
community members). In effect, this creates an echo chamber that is created by
community members and not platform algorithms. 
Finally, by demonstrating the effects of outright rejection, our work also
shows the responsibility that users have to be civil with each other. 

\para{Future research directions.}
The implications detailed above highlight some platform-level design and
administration strategies that could be adopted to prevent the spread of
extremism.
Our research suggests another interesting direction for future research: using 
observational textual data for the development of nuanced interventions for
the prevention of user radicalization.
Importantly, our methods for tracking changes in warning behaviors and
measuring events that influence these changes has the potential to help
us: (1) identify when a user might be in need of an intervention, (2) identify
when a user might be receptive to an intervention, and (3) understand the most
opportune time during which interventions might be applied to effectively
prevent the radicalization of vulnerable users by extremist groups.
In addition, the metrics used in this research for tracking warning behaviors
are usable as measures of the effectiveness of deradicalization strategies.

\balance

%\footnotesize

\bibliographystyle{unsrt}
\bibliography{reddit}

\begin{thebibliography}{10}

\bibitem{CapitolRiotsOnline}
Sheera Frenkel.
\newblock {How The Storming of Capitol Hill Was Organized on Social Media - The
  New York Times}.
\newblock
  \url{https://www.nytimes.com/2021/01/06/us/politics/protesters-storm-capitol-hill-building.html}.

\bibitem{UTROnline}
Alex Heath.
\newblock {Facebook Removed Unite the Right Charlottesville Rally Event Page
  One Day Before - Business Insider}.
\newblock
  \url{https://www.businessinsider.com/facebook-removed-unite-the-right-charlottesville-rally-event-page-one-day-before-2017-8}.

\bibitem{AMOnline}
Leyland Cecco.
\newblock {Toronto van attack suspect says he was radicalized online by incels
  - The Guardian}.
\newblock
  \url{https://www.theguardian.com/world/2019/sep/27/alek-minassian-toronto-van-attack-interview-incels}.

\bibitem{Weimann-BJWA2010}
Gabriel Weimann.
\newblock Terror on facebook, twitter, and youtube.
\newblock {\em The Brown Journal of World Affairs}, 16(2):45--54, 2010.

\bibitem{TerrorismFBI}
{Terrorism - FBI}.
\newblock \url{https://www.fbi.gov/investigate/terrorism}.

\bibitem{jointreport}
{2021 Strategic Intelligence Assessment and Data on Domestic Terrorism - FBI
  and DHS}.
\newblock
  \url{https://www.fbi.gov/file-repository/fbi-dhs-domestic-terrorism-strategic-report.pdf/view}.

\bibitem{massanari2017gamergate}
Adrienne Massanari.
\newblock \# gamergate and the fappening: How reddit's algorithm, governance,
  and culture support toxic technocultures.
\newblock {\em New media \& society}, 19(3):329--346, 2017.

\bibitem{farrell2019exploring}
Tracie Farrell, Miriam Fernandez, Jakub Novotny, and Harith Alani.
\newblock Exploring misogyny across the manosphere in reddit.
\newblock In {\em Proceedings of the 10th ACM Conference on Web Science}, pages
  87--96, 2019.

\bibitem{marwick2018drinking}
Alice~E Marwick and Robyn Caplan.
\newblock Drinking male tears: Language, the manosphere, and networked
  harassment.
\newblock {\em Feminist Media Studies}, 18(4):543--559, 2018.

\bibitem{OMalley-JIV2020}
Roberta~Liggett O'Malley, Karen Holt, and Thomas~J. Holt.
\newblock An exploration of the involuntary celibate (incel) subculture online.
\newblock {\em Journal of Interpersonal Violence}, 2020.
\newblock PMID: 32969306.

\bibitem{wright2020pussy}
Scott Wright, Verity Trott, and Callum Jones.
\newblock `the pussy ain't worth it, bro': assessing the discourse and
  structure of mgtow.
\newblock {\em Information, Communication \& Society}, 23(6):908--925, 2020.

\bibitem{news-pua-rape}
{Pick up artist raped woman and blogged about the attack - The Independent}.
\newblock
  \url{https://www.independent.co.uk/news/world/americas/pick-up-artist-jailed-after-raping-woman-and-blogging-about-the-attack-a7473831.html},
  Dec 2016.

\bibitem{news-pua-shooter}
Tracy Clark-Flory.
\newblock {Pickup artists: Gym shooter is one of us - Salon}.
\newblock \url{https://www.salon.com/2009/08/06/hatred_2/}, Sep 2011.

\bibitem{van2021digesting}
Shawn~P Van~Valkenburgh.
\newblock Digesting the red pill: Masculinity and neoliberalism in the
  manosphere.
\newblock {\em Men and Masculinities}, 24(1):84--103, 2021.

\bibitem{splc-mensrights}
Misogyny: The sites, Jan 1970.

\bibitem{ManosphereViolence}
{The misogynist incel movement is spreading. Should it be classified as a
  terror threat? - The Guardian}.
\newblock
  \url{https://www.theguardian.com/lifeandstyle/2021/mar/03/incel-movement-terror-threat-canada}.

\bibitem{tang2021down}
Lu~Tang, Kayo Fujimoto, Muhammad~Tuan Amith, Rachel Cunningham, Rebecca~A
  Costantini, Felicia York, Grace Xiong, Julie~A Boom, and Cui Tao.
\newblock ``down the rabbit hole'' of vaccine misinformation on youtube:
  Network exposure study.
\newblock {\em Journal of Medical Internet Research}, 23(1):e23262, 2021.

\bibitem{roose2019making}
Kevin Roose.
\newblock The making of a youtube radical.
\newblock {\em The New York Times}, 8, 2019.

\bibitem{faris2017partisanship}
Robert Faris, Hal Roberts, Bruce Etling, Nikki Bourassa, Ethan Zuckerman, and
  Yochai Benkler.
\newblock Partisanship, propaganda, and disinformation: Online media and the
  2016 us presidential election.
\newblock {\em Berkman Klein Center Research Publication}, 6, 2017.

\bibitem{bruns2021coronavirus}
Axel Bruns, Stephen Harrington, and Edward Hurcombe.
\newblock Coronavirus conspiracy theories: Tracing misinformation trajectories
  from the fringes to the mainstream.
\newblock In {\em Communicating COVID-19}, pages 229--249. Springer, 2021.

\bibitem{koehler2014radical}
Daniel Koehler.
\newblock The radical online: Individual radicalization processes and the role
  of the internet.
\newblock {\em Journal for Deradicalization}, (1):116--134, 2014.

\bibitem{kadivar2017online}
Jamileh Kadivar.
\newblock Online radicalization and social media: A case study of daesh.
\newblock {\em International Journal of Digital Television}, 8(3):403--422,
  2017.

\bibitem{blaker2015islamic}
Lisa Blaker.
\newblock The islamic state’s use of online social media.
\newblock {\em Military Cyber Affairs}, 1(1):4, 2015.

\bibitem{gaspar2020radicalization}
Hande~Abay Gaspar, Christopher Daase, Nicole Deitelhoff, Julian Junk, and
  Manjana Sold.
\newblock Radicalization and political violence--challenges of conceptualizing
  and researching origins, processes and politics of illiberal beliefs.
\newblock {\em International Journal of Conflict and Violence (IJCV)},
  14:1--18, 2020.

\bibitem{reidy2018radicalization}
Ken Reidy.
\newblock Radicalization as a vector: Exploring non-violent and benevolent
  processes of radicalization.
\newblock {\em Journal for Deradicalization}, (14):249--294, 2018.

\bibitem{smith2018radicalization}
Allison~G Smith.
\newblock {\em How radicalization to terrorism occurs in the United States:
  What research sponsored by the National Institute of Justice tells us}.
\newblock US Department of Justice, Office of Justice Programs, National
  Institute of Justice, 2018.

\bibitem{gill2015lone}
Paul Gill.
\newblock {\em Lone-actor terrorists: A behavioural analysis}.
\newblock Routledge, 2015.

\bibitem{meloywarning}
J~Reid~Meloy, Jens Hoffmann, Angela Guldimann, and David James.
\newblock The role of warning behaviors in threat assessment: An exploration
  and suggested typology.
\newblock {\em Behavioral sciences \& the law}, 30(3):256--279, 2012.

\bibitem{elster1996rationality}
Jon Elster.
\newblock Rationality and the emotions.
\newblock {\em The economic journal}, 106(438):1386--1397, 1996.

\bibitem{baele2017lone}
Stephane~J Baele.
\newblock Lone-actor terrorists’ emotions and cognition: An evaluation beyond
  stereotypes.
\newblock {\em Political Psychology}, 38(3):449--468, 2017.

\bibitem{doosje2013determinants}
Bertjan Doosje, Annemarie Loseman, and Kees Van Den~Bos.
\newblock Determinants of radicalization of islamic youth in the netherlands:
  Personal uncertainty, perceived injustice, and perceived group threat.
\newblock {\em Journal of Social Issues}, 69(3):586--604, 2013.

\bibitem{grover2019detecting}
Ted Grover and Gloria Mark.
\newblock Detecting potential warning behaviors of ideological radicalization
  in an alt-right subreddit.
\newblock In {\em Proceedings of the International AAAI Conference on Web and
  Social Media}, volume~13, pages 193--204, 2019.

\bibitem{cohen2014detecting}
Katie Cohen, Fredrik Johansson, Lisa Kaati, and Jonas~Clausen Mork.
\newblock Detecting linguistic markers for radical violence in social media.
\newblock {\em Terrorism and Political Violence}, 26(1):246--256, 2014.

\bibitem{kaati2016linguistic}
Lisa Kaati, Amendra Shrestha, and Katie Cohen.
\newblock Linguistic analysis of lone offender manifestos.
\newblock In {\em 2016 IEEE international conference on cybercrime and computer
  forensic (ICCCF)}, pages 1--8. IEEE, 2016.

\bibitem{assessments-dhs2018}
{Department of Homeland Security - The Application of Risk Assessment Tools in
  the Criminal Justice and Rehabilitation Process: Literature Review}.
\newblock
  \url{https://www.dhs.gov/sites/default/files/publications/OPSR_TP_CVE-Application-Risk-Assessment-Tools-Criminal-Rehab-Process_2018Feb-508.pdf}.

\bibitem{meloy2018operational}
J~Reid Meloy.
\newblock The operational development and empirical testing of the terrorist
  radicalization assessment protocol (trap--18).
\newblock {\em Journal of personality assessment}, 100(5):483--492, 2018.

\bibitem{pressman2012calibrating}
D~Elaine Pressman and John Flockton.
\newblock Calibrating risk for violent political extremists and terrorists: The
  vera 2 structured assessment.
\newblock {\em The British Journal of Forensic Practice}, 2012.

\bibitem{lloyd2019extremist}
Monica Lloyd.
\newblock Extremist risk assessments: a directory.
\newblock 2019.

\bibitem{lloyd2015development}
Monica Lloyd and Christopher Dean.
\newblock The development of structured guidelines for assessing risk in
  extremist offenders.
\newblock {\em Journal of Threat Assessment and Management}, 2(1):40, 2015.

\bibitem{cole2010guidance}
Jon Cole, Emily Alison, Ben Cole, and Laurence Alison.
\newblock Guidance for identifying people vulnerable to recruitment into
  violent extremism.
\newblock {\em Liverpool, UK: University of Liverpool, School of Psychology},
  2010.

\bibitem{van2021grievance}
Isabelle van~der Vegt, Maximilian Mozes, Bennett Kleinberg, and Paul Gill.
\newblock The grievance dictionary: understanding threatening language use.
\newblock {\em Behavior research methods}, pages 1--15, 2021.

\bibitem{smith2018risk}
Allison~G Smith.
\newblock {\em Risk factors and indicators associated with radicalization to
  terrorism in the United States: What research sponsored by the National
  Institute of Justice tells us}.
\newblock US Department Of Justice, Office of Justice Programs, National
  Institute of Justice, 2018.

\bibitem{mccauley2008mechanisms}
Clark McCauley and Sophia Moskalenko.
\newblock Mechanisms of political radicalization: Pathways toward terrorism.
\newblock {\em Terrorism and political violence}, 20(3):415--433, 2008.

\bibitem{corner2018mental}
Emily Corner, Paul Gill, Ronald Schouten, and Frank Farnham.
\newblock Mental disorders, personality traits, and grievance-fueled targeted
  violence: the evidence base and implications for research and practice.
\newblock {\em Journal of personality assessment}, 100(5):459--470, 2018.

\bibitem{smith2004words}
Allison~G Smith.
\newblock From words to action: Exploring the relationship between a group's
  value references and its likelihood of engaging in terrorism.
\newblock {\em Studies in Conflict \& Terrorism}, 27(5):409--437, 2004.

\bibitem{chung2011using}
Cindy~K Chung and James~W Pennebaker.
\newblock Using computerized text analysis to assess threatening communications
  and behavior.
\newblock {\em Threatening communications and behavior: Perspectives on the
  pursuit of public figures}, pages 3--32, 2011.

\bibitem{pressman2019internet}
D~Elaine Pressman and Cristina Ivan.
\newblock Internet use and violent extremism: A cyber-vera risk assessment
  protocol.
\newblock In {\em Violent Extremism: Breakthroughs in Research and Practice},
  pages 43--61. IGI Global, 2019.

\bibitem{stankov2010contemporary}
Lazar Stankov, Derrick Higgins, Gerard Saucier, and Goran Kne{\v{z}}evi{\'c}.
\newblock Contemporary militant extremism: A linguistic approach to scale
  development.
\newblock {\em Psychological assessment}, 22(2):246, 2010.

\bibitem{douglas1999assessing}
Kevin~S Douglas, James~RP Ogloff, Tonia~L Nicholls, and Isabel Grant.
\newblock Assessing risk for violence among psychiatric patients: the hcr-20
  violence risk assessment scheme and the psychopathy checklist: Screening
  version.
\newblock {\em Journal of consulting and clinical psychology}, 67(6):917, 1999.

\bibitem{sarma2017risk}
Kiran~M Sarma.
\newblock Risk assessment and the prevention of radicalization from nonviolence
  into terrorism.
\newblock {\em American Psychologist}, 72(3):278, 2017.

\bibitem{egan2016can}
Vincent Egan, Jon Cole, Ben Cole, Laurence Alison, Emily Alison, Sara Waring,
  and Stamatis Elntib.
\newblock Can you identify violent extremists using a screening checklist and
  open-source intelligence alone?
\newblock {\em Journal of Threat Assessment and Management}, 3(1):21, 2016.

\bibitem{berger2017extremist}
JM~Berger.
\newblock Extremist construction of identity.
\newblock {\em How Escalating Demands for Legitimacy Shape and Define In-Group
  and Out-Group Dybamics, ICCT Research Paper April}, 2017.

\bibitem{scrivens2021examining}
Ryan Scrivens, Amanda~Isabel Osuna, Steven~M Chermak, Michael~A Whitney, and
  Richard Frank.
\newblock Examining online indicators of extremism in violent right-wing
  extremist forums.
\newblock {\em Studies in Conflict \& Terrorism}, pages 1--25, 2021.

\bibitem{berbrier2000victim}
Mitch Berbrier.
\newblock The victim ideology of white supremacists and white separatists in
  the united states.
\newblock {\em Sociological Focus}, 33(2):175--191, 2000.

\bibitem{scrivens2018searching}
Ryan Scrivens, Garth Davies, and Richard Frank.
\newblock Searching for signs of extremism on the web: an introduction to
  sentiment-based identification of radical authors.
\newblock {\em Behavioral sciences of terrorism and political aggression},
  10(1):39--59, 2018.

\bibitem{wilson2018comparative}
Rebecca Wilson.
\newblock A comparative analysis of the implicit motives of violent extremist
  groups.
\newblock 2018.

\bibitem{tausczik2010psychological}
Yla~R Tausczik and James~W Pennebaker.
\newblock The psychological meaning of words: Liwc and computerized text
  analysis methods.
\newblock {\em Journal of language and social psychology}, 29(1):24--54, 2010.

\bibitem{hutto2014vader}
Clayton Hutto and Eric Gilbert.
\newblock Vader: A parsimonious rule-based model for sentiment analysis of
  social media text.
\newblock In {\em Proceedings of the International AAAI Conference on Web and
  Social Media}, volume~8, 2014.

\bibitem{wahlstrom2020dynamics}
Mattias Wahlstr{\"o}m, Anton T{\"o}rnberg, and Hans Ekbrand.
\newblock Dynamics of violent and dehumanizing rhetoric in far-right social
  media.
\newblock {\em New Media \& Society}, page 1461444820952795, 2020.

\bibitem{piazza2020politician}
James~A Piazza.
\newblock Politician hate speech and domestic terrorism.
\newblock {\em International Interactions}, 46(3):431--453, 2020.

\bibitem{krendel2020men}
Alexandra Krendel.
\newblock The men and women, guys and girls of the ‘manosphere’: A
  corpus-assisted discourse approach.
\newblock {\em Discourse \& Society}, 31(6):607--630, 2020.

\bibitem{Perspective-API-reddit}
Perspective api use cases.
\newblock \url{https://www.perspectiveapi.com/case-studies/}.

\bibitem{pavlopoulos2019convai}
John Pavlopoulos, Nithum Thain, Lucas Dixon, and Ion Androutsopoulos.
\newblock Convai at semeval-2019 task 6: Offensive language identification and
  categorization with perspective and bert.
\newblock In {\em Proceedings of the 13th international Workshop on Semantic
  Evaluation}, pages 571--576, 2019.

\bibitem{obadimu2021developing}
Adewale Obadimu, Tuja Khaund, Esther Mead, Thomas Marcoux, and Nitin Agarwal.
\newblock Developing a socio-computational approach to examine toxicity
  propagation and regulation in covid-19 discourse on youtube.
\newblock {\em Information Processing \& Management}, page 102660, 2021.

\bibitem{jaki2019online}
Sylvia Jaki, Tom De~Smedt, Maja Gw{\'o}{\'z}d{\'z}, Rudresh Panchal, Alexander
  Rossa, and Guy De~Pauw.
\newblock Online hatred of women in the incels. me forum: Linguistic analysis
  and automatic detection.
\newblock {\em Journal of Language Aggression and Conflict}, 7(2):240--268,
  2019.

\bibitem{bou2014gender}
Patricia Bou-Franch and Pilar Garc{\'e}s-Conejos Blitvich.
\newblock Gender ideology and social identity processes in online language
  aggression against women.
\newblock {\em Journal of Language Aggression and Conflict}, 2(2):226--248,
  2014.

\bibitem{ribero2021evolution}
Manoel Horta~Ribeiro, Jeremy Blackburn, Barry Bradlyn, Emiliano De~Cristofaro,
  Gianluca Stringhini, Summer Long, Stephanie Greenberg, and Savvas Zannettou.
\newblock The evolution of the manosphere across the web.
\newblock {\em Proceedings of the International AAAI Conference on Web and
  Social Media}, 2021.

\bibitem{mamie2021anti}
Robin Mami{\'e}, Manoel~Horta Ribeiro, and Robert West.
\newblock Are anti-feminist communities gateways to the far right? evidence
  from reddit and youtube.
\newblock {\em arXiv:2102.12837}, 2021.

\bibitem{habib2019act}
Hussam Habib, Maaz~Bin Musa, Fareed Zaffar, and Rishab Nithyanand.
\newblock To act or react: Investigating proactive strategies for online
  community moderation.
\newblock {\em arXiv preprint arXiv:1906.11932}, 2019.

\bibitem{habib2021proactive}
Hussam Habib, Maaz~Bin Musa, Fareed Zaffar, and Rishab Nithyanand.
\newblock Are proactive interventions for reddit communities feasible?
\newblock {\em Proceedings of the International AAAI Conference on Web and
  Social Media}, 2022.

\bibitem{hine2017kek}
Gabriel Hine, Jeremiah Onaolapo, Emiliano De~Cristofaro, Nicolas Kourtellis,
  Ilias Leontiadis, Riginos Samaras, Gianluca Stringhini, and Jeremy Blackburn.
\newblock Kek, cucks, and god emperor trump: A measurement study of 4chan’s
  politically incorrect forum and its effects on the web.
\newblock In {\em Proceedings of the International AAAI Conference on Web and
  Social Media}, 2017.

\bibitem{zannettou2017web}
Savvas Zannettou, Tristan Caulfield, Emiliano De~Cristofaro, Nicolas
  Kourtelris, Ilias Leontiadis, Michael Sirivianos, Gianluca Stringhini, and
  Jeremy Blackburn.
\newblock The web centipede: understanding how web communities influence each
  other through the lens of mainstream and alternative news sources.
\newblock In {\em Proceedings of the 2017 internet measurement conference},
  2017.

\bibitem{We-are-the-nerds}
Christine Lagorio-Chafkin.
\newblock {\em We Are the Nerds: The Birth and Tumultuous Life of Reddit}.
\newblock Hachette, 2018.

\bibitem{PushShift-Reddit}
Jason Baumgartner, Savvas Zannettou, Brian Keegan, Megan Squire, and Jeremy
  Blackburn.
\newblock The pushshift reddit dataset.
\newblock In {\em Proceedings of the International AAAI Conference on Web and
  Social Media}, 2020.

\bibitem{mikolov2013efficient}
Tomas Mikolov, Kai Chen, Greg Corrado, and Jeffrey Dean.
\newblock Efficient estimation of word representations in vector space, 2013.

\bibitem{devlin2019bert}
Jacob Devlin, Ming-Wei Chang, Kenton Lee, and Kristina Toutanova.
\newblock Bert: Pre-training of deep bidirectional transformers for language
  understanding, 2019.

\bibitem{clustermetric}
Peter~J. Rousseeuw.
\newblock Silhouettes: A graphical aid to the interpretation and validation of
  cluster analysis.
\newblock {\em Journal of Computational and Applied Mathematics}, 20:53--65,
  1987.

\bibitem{Manosphere-subreddits}
{iDrama Lab: Manosphere Analysis - GitHub}.
\newblock
  \url{https://github.com/idramalab/manosphere_analysis/blob/master/data/subreddit_descriptions.csv}.

\bibitem{muller2007dynamic}
Meinard M{\"u}ller.
\newblock Dynamic time warping.
\newblock {\em Information retrieval for music and motion}, pages 69--84, 2007.

\bibitem{molnar2020interpretable}
Christoph Molnar.
\newblock {\em Interpretable machine learning}.
\newblock Lulu. com, 2020.

\bibitem{phadke2021makes}
Shruti Phadke, Mattia Samory, and Tanushree Mitra.
\newblock What makes people join conspiracy communities? role of social factors
  in conspiracy engagement.
\newblock {\em Proceedings of the ACM on Human-Computer Interaction},
  4(CSCW3):1--30, 2021.

\bibitem{wong2015supremacy}
Meghan~A Wong, Richard Frank, and Russell Allsup.
\newblock The supremacy of online white supremacists--an analysis of online
  discussions by white supremacists.
\newblock {\em Information \& Communications Technology Law}, 24(1):41--73,
  2015.

\bibitem{evans2018memes}
R~Evans.
\newblock From memes to infowars: How 75 fascist activists were red-pilled.
\newblock {\em Bell?` ngcat, October}, 2018.

\bibitem{renstrom2020exploring}
Emma~A Renstr{\"o}m, Hanna B{\"a}ck, and Holly~M Knapton.
\newblock Exploring a pathway to radicalization: The effects of social
  exclusion and rejection sensitivity.
\newblock {\em Group Processes \& Intergroup Relations}, 23(8):1204--1229,
  2020.

\bibitem{haddad2021online}
Hanin Haddad, Nisha Baral, and Ivan Garibay.
\newblock Online rejection influence on behavior deviancy and radicalization:
  An agent-based model approach.
\newblock In {\em Proceedings of the 2020 Conference of The Computational
  Social Science Society of the Americas}, pages 15--29. Springer, 2021.

\bibitem{jasko2017quest}
Katarzyna Jasko, Gary LaFree, and Arie Kruglanski.
\newblock Quest for significance and violent extremism: The case of domestic
  radicalization.
\newblock {\em Political Psychology}, 38(5):815--831, 2017.

\bibitem{scrivens2015sentiment}
Ryan Scrivens, Garth Davies, Richard Frank, and Joseph Mei.
\newblock Sentiment-based identification of radical authors (sira).
\newblock In {\em 2015 IEEE International Conference on Data Mining Workshop
  (ICDMW)}, pages 979--986. IEEE, 2015.

\bibitem{ribeiro2020auditing}
Manoel~Horta Ribeiro, Raphael Ottoni, Robert West, Virg{\'\i}lio~AF Almeida,
  and Wagner Meira~Jr.
\newblock Auditing radicalization pathways on youtube.
\newblock In {\em Proceedings of the 2020 conference on fairness,
  accountability, and transparency}, pages 131--141, 2020.

\bibitem{nguyen2020echo}
C~Thi Nguyen.
\newblock Echo chambers and epistemic bubbles.
\newblock {\em Episteme}, 17(2):141--161, 2020.

\bibitem{chitra2020analyzing}
Uthsav Chitra and Christopher Musco.
\newblock Analyzing the impact of filter bubbles on social network
  polarization.
\newblock In {\em Proceedings of the 13th International Conference on Web
  Search and Data Mining}, pages 115--123, 2020.

\bibitem{farrell2020use}
Tracie Farrell, Oscar Araque, Miriam Fernandez, and Harith Alani.
\newblock On the use of jargon and word embeddings to explore subculture within
  the reddit's manosphere.
\newblock In {\em 12th ACM Conference on Web Science}, pages 221--230, 2020.

\end{thebibliography}

%\newpage

%\input{appendix}

\end{document}